\documentclass[aps,pra,twocolumn]{revtex4}
\usepackage[dvips]{epsfig}
\usepackage[usenames]{color}
\usepackage{pstricks}
\usepackage{amsmath}
\usepackage{amssymb}
\usepackage{subfigure}
\usepackage{afterpage}
\usepackage{nameref}
\usepackage{array}
\usepackage{multirow}

\begin{document}

\title{Dynamics of ultracold dipolar particles in a confined geometry and tilted fields}

\author{Goulven Qu{\'e}m{\'e}ner, Maxence Lepers, Olivier Dulieu}
\affiliation{Laboratoire Aim{\'e} Cotton, CNRS, Universit{\'e} Paris-Sud, ENS Cachan,
91405 Orsay, France}

\date{\today}

\begin{abstract}
We develop a collisional formalism adapted for the dynamics of ultracold dipolar particles in 
a confined geometry and in fields tilted relative to the confinement axis. 
Using tesseral harmonics instead of the usual 
spherical harmonics to expand the scattering wavefunction, we recover a good quantum number  
$\xi = \pm 1$ which is conserved during the collision. We derive the general expression of 
the dipole-dipole interaction in this convenient basis set as a function of the polar 
and azimuthal angles of the fields. We apply the formalism to the collision of fermionic and bosonic
polar KRb molecules in a tilted electric field and in a one-dimensional optical lattice.
The presence of a tilted field drastically changes  
the magnitude of the reactive and inelastic rates
as well as the inelastic threshold properties at vanishing collision energies. Setting an appropriate 
strength of the 
confinement for the fermionic system, we show that the ultracold particles can even further reduce 
their kinetic energy by inelastic excitation to higher states of the confinement trap.
\end{abstract}

\maketitle

\font\smallfont=cmr7

\section{Introduction}

The field of ultracold gases composed of dipolar particles 
has generated tremendous interest during the past
years~\cite{Doyle_EPJD_31_149_2004,Baranov_PRep_464_71_2008,
Lahaye_RPP_72_126401_2009,
Quemener_CR_112_4949_2012,Baranov_CR_112_5012_2012,
Kotochigova_RPP_77_093901_2014}. 
One major goal 
is to shape at will the quantum properties
of ultracold gases using the high degree of controllability available in experiments.
Different kinds of dipolar particles are concerned by the quest for manifestation of dipole-induced effects. 
The first category of interest contains particles with electric dipoles such as ground state molecules of KRb~\cite{Ni_S_322_231_2008,Aikawa_PRL_105_203001_2010}, RbCs~\cite{Takekoshi_PRL_113_205301_2014,Molony_PRL_113_255301_2014}, 
NaK~\cite{Park_PRL_114_205302_2015} and many others experimentally under way... 
The second category includes particles with magnetic dipoles such as Cr~\cite{Griesmaier_PRL_94_160401_2005,
Beaufils_PRA_77_061601_2008,Naylor_PRA_91_011603_2015},
Dy~\cite{Lu_PRL_107_190401_2011,Lu_PRL_108_215301_2012}, Er~\cite{Aikawa_PRL_108_210401_2012,Aikawa_PRL_112_010404_2014} atoms and Er$_2$ molecules~\cite{Frisch_arXiv_1504_04578_2015}...
The third category consists of particles with both electric and magnetic dipoles such as molecules of OH~\cite{Stuhl_N_492_396_2012}, SrF~\cite{Barry_N_512_286_2014}, YO~\cite{Yeo_PRL_114_223003_2015,Collopy_NJP_17_055008_2015}, RbSr~\cite{Pasquiou_PRA_88_023601_2013}...
All these particles can be manipulated by different configurations of electric and/or magnetic fields. 
They can also be loaded in optical lattices of different dimensions
such as one dimensional (1D) lattices \cite{Ticknor_PRA_81_042708_2010,
Quemener_PRA_81_060701_2010,Micheli_PRL_105_073202_2010,
Quemener_PRA_83_012705_2011,DeMiranda_NP_7_502_2011},
two dimensional (2D) lattices~\cite{Chotia_PRL_108_080405_2012,Simoni_NJP_17_013020_2015}
and three dimensional (3D) lattices~\cite{Yan_N_501_521_2013}. 
Due to the wide and numerous domains of application of dipolar 
particles~\cite{Carr_NJP_11_055049_2009},
it is therefore important to understand how to control 
the interactions and the collisional properties 
of these particles under such configurations of fields and lattices.
For example, it has been shown that chemical reactivity of molecules can be suppressed
by an electric field in a confined 1D optical lattice~\cite{DeMiranda_NP_7_502_2011} 
or by selecting a particular electric field and appropriate quantum states 
of the molecules in a non-confined space~\cite{Wang_NJP_17_035015_2015}.

\begin{figure}[h]
\begin{center}
\includegraphics[width=0.3\textwidth,keepaspectratio=true,angle=-90,trim=1cm 1.5cm 9cm 0cm,clip=true]{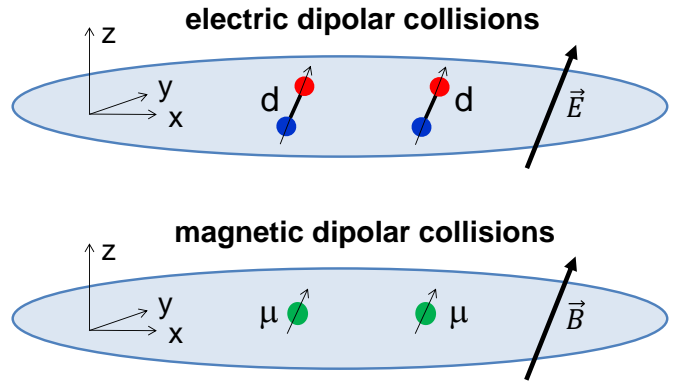} 
\end{center}
\caption{(Color online) Electric and magnetic dipolar collisions in a 1D confinement (pancakes-shaped lattices) and arbitrary tilted fields relative to the $z$ and $x$ axis. We consider ``classical'' dipoles aligned with the fields (see text).}
\label{DD2DTILT-1-FIG}
\end{figure}

In this study, we investigate two-body collisions in a confining 1D optical lattice in electric and/or magnetic fields tilted
with respect to the $z$ and $x$ axis, as illustrated in Fig.~\ref{DD2DTILT-1-FIG}.
We consider here ``classical'' dipoles 
aligned along the field, for which
the angular internal structure
of the particles (rotational angular momentum for the electric dipoles
and electronic angular momentum for the magnetic dipoles) is not taken into account. 
We choose an effective 
value $d$ or $\mu$ of the electric or magnetic dipole moment, corresponding to their 
expectation value along the direction of the electric 
or magnetic field.
For particles without angular internal structure, the total angular momentum $\hat{J}$ 
of the two-body colliding system reduces to the orbital angular momentum $\hat{L}$ associated with the quantum number $l$, 
which is conserved in free space.
When a field is applied parallel to the quantization axis $z$, 
it is not conserved anymore
but its projection $\hat{L}_z$ associated with the quantum number $m_l$ still is. 
When the field is tilted and no more parallel with respect to the quantization axis, its projection is not conserved anymore. The 
scattering problem becomes challenging as $l,m_l$ 
are all 
mixed.  
We show in this study that it is still possible to define a good quantum number,
provided that the collision is described using the so-called ``tesseral harmonics''~\cite{Whittaker_Book_1990} 
instead of the standard spherical harmonics
for the partial wave expansion of the scattering wavefunction.
The problem is thus split in two sub-problems of smaller size.
As an example of a dipolar system, we study 
fermionic and bosonic KRb + KRb collisions in a tilted electric field.
The present theoretical formalism was also successfully used recently to understand the  
experimental observation of dipolar collisions of bosonic Feshbach Er$_2$ molecules in a 1D 
optical lattice in a tilted magnetic field~\cite{Frisch_arXiv_1504_04578_2015}.

The paper is structured as follows. In Section II, we develop the theoretical formalism for 
collision of particles in a tilted field and confined geometry where the appropriate basis set for partial wave expansion 
is introduced. In Section III, we show how the collisional rate coefficients and 
their threshold behaviours are strongly affected by tilted fields revealing the complexity of the 
mechanism. The role of the confinement trap is also explored and could be used to reduce the kinetic energy of the particles. Finally, we conclude in Section IV.

\section{Theoretical formalism}

Ultracold dipolar collisions in a tilted 
field have been already studied in the past including 
microwave fields~\cite{Avdeenkov_NJP_11_055016_2009,Avdeenkov_PRA_86_022707_2012,
Avdeenkov_NJP_17_045025_2015} but without confinement,
in crossed electric and magnetic fields~\cite{Abrahamsson_JCP_127_044302_2007,Quemener_PRA_88_012706_2013},
or considering strong 1D confinement such that the particles are bound 
to collide in pure 2D~\cite{Ticknor_PRA_84_032702_2011}.
By pure 2D we mean that the characteristic strength of the particles confinement
is much stronger than the characteristic strength of their interaction (the 
dipole-dipole interaction here). However, this regime of pure 2D 
collisions is not yet reached in 
ongoing experiments as the required confinement strength is too strong.
Instead, quasi-2D collisions occur. The particles start to collide at large distances in pure 2D 
but there is a point as they approach each other where
the increasing magnitude of their interaction gets much bigger than the confinement strength. 
The particles do not feel anymore the presence of a 2D confinement
and collide as if they were in a non-confined space.
Therefore to reproduce the conditions of ongoing experiments, 
we describe the quasi-2D collisions of two 
ultracold particles of mass $m_1,m_2$ carrying 
tilted dipole moments and trapped in a 1D optical lattice of arbitrary confinement strength.
We assume that the particles cannot hop from 
one potential well to another so that we approximate a well 
by a harmonic oscillator for particle 1 and 2, 
$V_{ho} = 1/2 \ m_1 \, \omega^2 \, z^2_1 + 1/2 \ m_2 \, \omega^2 \, z^2_2$.
The angular frequency $\omega = 2 \pi \nu$ governs the strength of the confinement. 
The particles are initially in a given state of the harmonic 
oscillator $n_1,n_2$ of energy 
$\varepsilon_{n_1}=h \nu (n_1+1/2), \varepsilon_{n_2} = h \nu (n_2+1/2) $.

\begin{figure}[h]
\begin{center}
\includegraphics*[totalheight=0.5\textheight,keepaspectratio=true,angle=-90,trim=2.2cm 1.5cm 9cm 0cm,clip=true]{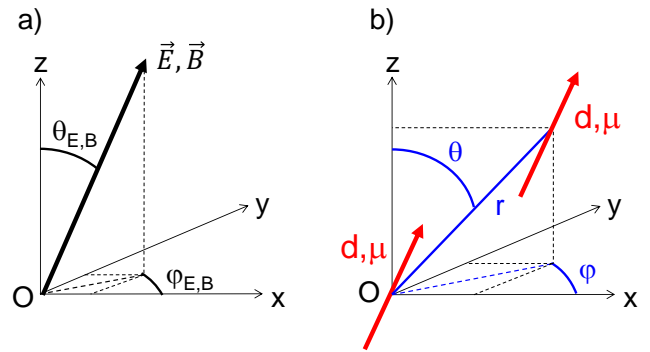} 
\end{center}
\caption{(Color online) a) Spherical angle coordinates $\theta_{E,B}$ relative to the $z$ axis and $\varphi_{E,B}$ relative to the $x$ axis of the tilted fields $\vec{E}$ and $\vec{B}$. In the study, the numerical results correspond to a field in the $xOz$ plane with $\varphi_{E,B}=0$.
b) Spherical coordinates $(r,\theta,\varphi)$ of the relative coordinate vector $\vec{r}$ describing the dipole-dipole collisional motion.}
\label{DD2DTILT-2-FIG}
\end{figure}

The electric or magnetic fields make an angle $\theta_{E,B}$ and $\varphi_{E,B}$
with the $z$ and $x$ axis respectively as depicted in Fig.~\ref{DD2DTILT-2-FIG}-a.
The classical dipole approximation has been shown convenient
for modelling electric dipoles~\cite{Wang_NJP_17_035015_2015} 
and magnetic dipoles~\cite{Frisch_arXiv_1504_04578_2015} 
interactions in ultracold dipolar gases. It is
an appropriate way to avoid the inclusion of the
particles internal structure in the collisional formalism, thus
sparing computational effort. No Stark nor Zeeman term
appears in our formalism then.

We distinguish three types of collisional processes:
(i) Elastic processes for 
which the molecules remain in the same external state of the harmonic oscillator after the collision: $(n'_1, n'_2) = (n_1, n_2)$; (ii): Inelastic processes for which the molecules change their external state of the harmonic oscillator $(n'_1, n'_2) \neq (n_1, n_2)$ 
(note that a change of the internal states is not possible as the internal structure of the particles is not treated); (iii) Loss processes due for instance to chemical reactions occurring at ultralow energy, leading to products
with high kinetic energy which are expelled from the trap.

\subsection{Quasi-2D collisions in parallel field}

We shall briefly recall the formalism for fields parallel to the quantization axis
when $\theta_{E,B}$=0. It is presented in more details in Ref.~\cite{Quemener_PRA_83_012705_2011}. 
First, it is convenient to transform the motion of the individual particles 1 and 2
with position coordinates $\vec{r}_1, \vec{r}_2$ of masses $m_1, m_2$ in external states $n_1, n_2$,
into the motion of two effective particles 
described by the center-of-mass and relative coordinates $\vec{R}, \vec{r}$
of masses $m_\text{tot}, m_\text{red}$ in external states $N, n$. 
For confinements modelled by harmonic oscillators, 
it can be shown then that the center-of-mass 
coordinate $\vec{R}$ is 
decoupled from the relative coordinate $\vec{r}$.

The potential energy term for the relative coordinate $\vec{r}$ 
is given by a van der Waals interaction
\begin{eqnarray}
V_{vdW} &=& -C_6/r^6 ,
\end{eqnarray}
the confinement interaction 
given by a harmonic oscillator for the relative motion along $z$
\begin{eqnarray}
V_{ho}  &=& \frac{1}{2} \, m_\text{red} \, \omega^2 \, z^2 ,
\end{eqnarray}
and a dipole-dipole interaction composed of an electric and magnetic term
\begin{eqnarray}
V_{dd}  &=&  V_{dd}^{e}  + V_{dd}^{m} 
\end{eqnarray}
where
\begin{eqnarray}
V_{dd}^{e,m}  = - X_{e,m}^2 (1 - 3 \cos^2\theta)/r^3 
\end{eqnarray}
with for electric dipoles 
\begin{eqnarray}
X_e \equiv d/\sqrt{4 \pi \varepsilon_0}
\end{eqnarray}
and for magnetic dipoles
\begin{eqnarray}
X_m \equiv \mu/\sqrt{(4 \pi/ \mu_0)} .
\end{eqnarray} 
The coordinate $\vec{r}$ is expressed in spherical coordinates 
$\vec{r} = (r, \theta, \varphi)$ when $V_{vdW}$ and $V_{dd}^{e,m}$
are dominant (Fig.~\ref{DD2DTILT-2-FIG}-b),
or in cylindrical coordinates $\vec{r} = (\rho, z, \varphi)$ when $V_{ho}$ is dominant 
at very large distances. 

We decompose the total wavefunction $\psi(\vec{r})$ of the relative motion 
into a basis set of spherical harmonics $Y_l^{m_l}(\theta, \varphi)\equiv \langle \hat{r}|l, m_l \rangle$ 
\begin{eqnarray}
\psi(r,\theta,\varphi) =   
\frac{1}{r} \, \sum_{l,m_l} f_{l,m_l}(r) \ Y_l^{m_l}(\theta, \varphi) 
\label{Psisph}
\end{eqnarray}
with $-l \le m_l \le +l$.
In this basis set, the expression of the potential energy terms become in bra-ket notation
\begin{eqnarray}
\langle l \, m_l | V_{vdW} | l' \, m_l' \rangle =  
- \frac{C_6}{r^6} \times  \delta_{m_l,m_l'} \, \delta_{l,l'},
\label{VVDW}
\end{eqnarray}
\begin{multline}
\langle l \, m_l | V_{ho} | l' \, m_l' \rangle = \delta_{m_l,m_l'} \\
 \frac{\frac{1}{2} \, m_\text{red} \, \omega^2 \, r^2}{3} \,
 \bigg\{  (-1)^{m_l}  \, \sqrt{2l+1} \, \sqrt{2l'+1}  \\ 
 \times 2 \,  \left( \begin{array}{ccc} l & 2 & l' \\ 0 & 0 & 0 \end{array} \right)
\, \left( \begin{array}{ccc} l & 2 & l' \\ -m_{l} & 0 & m_{l}' \end{array} \right) 
+ \delta_{l,l'}  \bigg\} ,
\label{VHO}
\end{multline}
and
\begin{multline}
\langle l \, m_l | V_{dd}^{e,m} | l' \, m_l' \rangle =   
 - \frac{\sqrt{30} \, X_{e,m}^2}{r^3} \\  
 \delta_{m_l,m_l'} \, (-1)^{m_l} \, \sqrt{2l+1} \, \sqrt{2l'+1} \\
   \left( \begin{array}{ccc} 1 & 1 & 2 \\ 0 & 0 & 0 \end{array} \right) 
\, \left( \begin{array}{ccc} l & 2 & l' \\ 0 & 0 & 0 \end{array} \right)
\, \left( \begin{array}{ccc} l & 2 & l' \\ -m_{l} & 0 & m_{l}' \end{array}  \right) .
\label{VDD}
\end{multline}
The first expression is diagonal in $m_l$ and $l$ while the two last expressions 
are diagonal in $m_l$ 
but couple different values of $l$.
The expression for $V_{ho}$ arises from the fact that 
$ z^2 = r^2 \, \cos^2 \theta = r^2 \, [4 \, \sqrt{\pi/5} \, Y^0_2 + 1] /3$.

The time-independent Schr{\"o}dinger equation is solved for a fixed total energy 
$E_\text{tot} = \varepsilon_{n_1} + \varepsilon_{n_2} + E_\text{coll}$. Equations \eqref{VVDW} to \eqref{VDD} lead to a set of coupled differential equations.
To solve this system of equations, 
we use a diabatic-by-sector method~\cite{Quemener_PRA_83_012705_2011} 
which generates a set of adiabatic energy curves 
as a function of the inter-particle distance $r$,
and we propagate the log-derivative of the wavefunction~\cite{Johnson_JCP_13_445_1973,Manolopoulos_JCP_85_6425_1986}.
The boundary condition at short distance, 
where the propagation of the wavefunction is 
started, is set up by a tunable, diagonal log-derivative matrix given in Ref.~\cite{Wang_NJP_17_035015_2015} for which we can control the amount of loss in this short-range region.
For the boundary condition at large distance 
in the asymptotic region where the propagation of 
the wavefunction is ended, we use the 
asymptotic form of the cylindrical wavefunction which is a linear combination of regular and
irregular Bessel functions of the cylindrical problem. Using a  transformation
matrix from cylindrical to spherical coordinates~\cite{Quemener_PRA_83_012705_2011}, 
we obtain the asymptotic form of the wavefunction
from which we deduce the $K$, $S$, and $T$ matrices using the expression of the 
log-derivative matrix 
in spherical coordinates. The $S$ matrix in the center-of-mass 
and relative coordinates is then expressed back into the individual
coordinates~\cite{Quemener_PRA_83_012705_2011} yielding the elastic, inelastic and loss cross 
sections and rate coefficients for two particles starting in a given initial state $(n_1,n_2)$.

\subsection{Quasi-2D collisions in an arbitrary field}

\subsubsection*{In the spherical harmonics basis set}

As depicted in Fig.~\ref{DD2DTILT-2-FIG}, when the fields are tilted, 
the expression of the dipole-dipole interactions become
\begin{eqnarray}
V_{dd}^{e,m} = - X_{e,m}^2 (1 - 3 \cos^2[\theta-\theta_{E,B}])/r^3 .
\end{eqnarray}
This implies a more complicated expression in the spherical harmonic basis set
\begin{multline}
\langle l \, m_l | V_{dd}^{e,m} | l' \, m_l' \rangle =  
 - \frac{\sqrt{30} \, X_{e,m}^2}{r^3} \\
 (-1)^{m_l} \, \sqrt{2l+1} \, \sqrt{2l'+1} \\
\shoveleft{ 
\sum_{p_1=-1}^{1} \, \sum_{p_2=-1}^{1} \, \sum_{p=-2}^{2} 
\, \left( \begin{array}{ccc} 1 & 1 & 2 \\ p_1 & p_2 & -p \end{array} \right) } \\
 \sqrt{4 \pi/3} \, Y_1^{p_1}(\theta_{E,B},\varphi_{E,B}) 
\, \sqrt{4 \pi/3} \, Y_1^{p_2}(\theta_{E,B},\varphi_{E,B}) \\
 \left( \begin{array}{ccc} l & 2 & l' \\ 0 & 0 & 0 \end{array} \right)
\, \left( \begin{array}{ccc} l & 2 & l' \\ -m_{l} & -p & m_{l}' \end{array} \right) .
\label{TRIPLESUM}
\end{multline}
For the special case $\theta_{E,B}=0$, $p_1=p_2=p=0$ and we recover the case
$m_l' - m_l = 0$. For the special case $\theta_{E,B}=\pi/2$, $p_1=\pm1$, $p_2=\pm1$, $p=0,\pm2$ then we have $m_l' - m_l = 0, \pm 2$.
For any other $\theta_{E,B}$ we have $m_l' - m_l = 0, \pm1, \pm2$, increasing the number of coupled equations and leaving no good quantum numbers in Eq.~\eqref{TRIPLESUM}.

\begin{table*}[t]
\setlength{\extrarowheight}{2pt}
\begin{tabular}{|c|c|c||c|c|c||c|c|c||c|c|c|}
  \hline
  \multicolumn{3}{|c||}{$\theta_{E,B} = 0$, $\varphi_{E,B} = 0$}  & \multicolumn{3}{c||}{$\theta_{E,B} = \pi/2$, $\varphi_{E,B} = 0$} & \multicolumn{3}{c||}{$\theta_{E,B} \ne 0,\pi/2$, $\varphi_{E,B} = 0$} & \multicolumn{3}{c|}{$\varphi_{E,B} \ne 0$}
\\
  \hline
  \multirow{4}{*}{$\xi=+1$} &  $\bar{m}_l = 0$ & y & 
  \multirow{4}{*}{$\xi=+1$} & \multirow{2}{1.cm}{$\bar{m}_l = 0,2,...$} & \multirow{2}{*}{y} & 
  \multirow{4}{*}{$\xi=+1$} &  \multirow{4}{1.5cm}{$\bar{m}_l = 0,1,2,3,...$} & \multirow{4}{*}{y} & 
  \multirow{8}{*}{$\xi=\pm1$} & \multirow{8}{1.5cm}{$\bar{m}_l = 0,1,2,3,...$} & \multirow{8}{*}{y} \\
 & $\bar{m}_l = 1$ & y & & & & & & & & & \\ 
 & $\bar{m}_l = 2$ & y & & \multirow{2}{1.cm}{$\bar{m}_l = 1,3,...$} & \multirow{2}{*}{y} & & & & & &\\
 & $\bar{m}_l = ...$ & y & & & & & & & & & \\ 
\cline{1-9}
  \multirow{4}{*}{$\xi=-1$} &  $\bar{m}_l = 1$ & n & 
  \multirow{4}{*}{$\xi=-1$} & \multirow{2}{1.cm}{$\bar{m}_l = 2,4,...$} & \multirow{2}{*}{y} &
  \multirow{4}{*}{$\xi=-1$} & \multirow{4}{1.5cm}{$\bar{m}_l = 1,2,3,4...$} & \multirow{4}{*}{y} & & & \\
 & $\bar{m}_l = 2$ & n & & & & & & & & & \\ 
 & $\bar{m}_l = 3$ & n & & \multirow{2}{1.cm}{$\bar{m}_l = 1,3,...$} & \multirow{2}{*}{y} & & & & & & \\
 & $\bar{m}_l = ...$ & n & & & & & & & & & \\ 
 \hline
\end{tabular}
\caption{Quantum numbers $\xi,\bar{m}_l$ needed ($y$) or not ($n$) for different angle $\theta_{E,B},\varphi_{E,B}$ of the fields. Note that for identical and indistinguishable (same internal state) bosons, $l$ is even: $l=0,2,4,...$ For identical and indistinguishable fermions, $l$ is odd: $l=1,3,5,...$. For distinguishable  or non-identical particles, both parities of 
$l$ should be included: $l=0,2,4,...$ and $l=1,3,5,...$.
For the special cases $\theta_{E,B}=0$ and $\pi/2$, if we start with molecules in the ground state of the harmonic oscillator, $m_l$ should be odd for fermions and even for bosons~\cite{Quemener_PRA_83_012705_2011}.}
\label{TABLE1}
\end{table*}

\subsubsection*{In the tesseral harmonics basis set}

By properly symmetrizing the basis set of spherical harmonics, we can still 
recover a good quantum number. We use the following symmetrized spherical harmonics 
in ket notation
\begin{multline}
|l, \bar{m}_l, \xi \rangle = \frac{\delta_{1 \, \xi} + i \, \delta_{-1 \, \xi}}{\sqrt{2 \, \Delta}}  \\ 
\times \left\{ |l, - \, |m_l| \, \rangle + \xi \, (-1)^{|m_l|} \, |l, \, |m_l| \, \rangle \right\} \label{TESSHARMKET}
\end{multline}
where $\bar{m}_l \equiv |m_l|$ and 
$\Delta \equiv 1 + \delta_{\bar{m}_l,0}$. 
The new quantum number $\xi$ takes the values $\xi=\pm1$ ,while $\bar{m}_l = 0,1,2,3 ... $ when $\xi = +1$ 
and $\bar{m}_l=1,2,3 ...$ when $\xi = -1$.
This new basis set are often called the tesseral harmonics~\cite{Whittaker_Book_1990}. We note them
$Y_{l,{\bar{m}_l,\xi}}$ with
\begin{eqnarray}
Y_{l,{\bar{m}_l \ne 0,\xi=+1}} & = & \frac {1}{\sqrt2} \left\{ Y_l^{-\bar{m}_l} + (-1)^{\bar{m}_l} \, Y_l^{\bar{m}_l} \right\} \nonumber \\ 
& \propto & P_l^{\bar{m}_l}(\cos{\theta}) \, \cos{\bar{m}_l \varphi} \nonumber \\
Y_{l,{\bar{m}_l=0,\xi=+1}} & = & Y_l^{\bar{m}_l=0}  \propto P_l(\cos{\theta}) \nonumber \\
Y_{l,{\bar{m}_l \ne 0,\xi=-1}} & = & \frac {i}{\sqrt2} \left\{ Y_l^{-\bar{m}_l} - (-1)^{\bar{m}_l} \, Y_l^{\bar{m}_l} \right\} \nonumber \\
& \propto & P_l^{\bar{m}_l}(\cos{\theta}) \, \sin{\bar{m}_l \varphi} . \label{TESSHARM}
\end{eqnarray}
In this new basis set, the 
potential energy matrix elements are given by
\begin{eqnarray}
 \langle l, \bar{m}_l, \xi | V_{vdW} | l', \bar{m}_l', \xi' \rangle =  
- \frac{C_6}{r^6}  \times \delta_{\xi,\xi'} \, \delta_{\bar{m}_l,\bar{m}_l'} 
\, \delta_{l,l'}  , \label{VVDWSYM}
\end{eqnarray}
\begin{multline}
\langle l, \bar{m}_l, \xi | V_{ho} | l', \bar{m}_l', \xi' \rangle =  
  \frac{1}{\sqrt{\Delta \, \Delta'}} \, \delta_{\xi,\xi'} \, 
  \delta_{\bar{m}_l,\bar{m}_l'} \\
 \frac{\frac{1}{2} \, m_\text{red} \, \omega^2 \, r^2}{3} \,
 \bigg\{  (-1)^{m_l}  \, \sqrt{2l+1} \, \sqrt{2l'+1}  \\ 
 \times 2 \,  \left( \begin{array}{ccc} l & 2 & l' \\ 0 & 0 & 0 \end{array} \right)
  \bigg[ 
   \left( \begin{array}{ccc} l & 2 & l' \\ \bar{m}_l & 0 & -\bar{m}_l' \end{array} \right) \\
 + (-1)^{\bar{m}_l'} \, \xi' \, \left( \begin{array}{ccc} l & 2 & l' \\ \bar{m}_l & 0 & \bar{m}_l' \end{array} \right) \, \delta_{\bar{m}_l,0} \, 
 \delta_{\xi',1} \,
        \bigg] \\        
+  
  \delta_{l,l'} + \delta_{l,l'} \, \xi' \, (-1)^{\bar{m}_l'} \, \delta_{\bar{m}_l,0} \, 
 \delta_{\xi',1} 
         \bigg\} , 
\label{VHOSYM}
\end{multline}
and after reducing Eq.~\eqref{TRIPLESUM}
\begin{widetext}
\begin{multline}
\langle l, \bar{m}_l, \xi | V_{dd}^{e,m} | l', \bar{m}_l', \xi' \rangle =  
 - \frac{\sqrt{30} \, X_{e,m}^2}{r^3} \, \frac{1}{\sqrt{\Delta \, \Delta'}} 
 (-1)^{\bar{m}_l} \, \sqrt{2l+1} \, \sqrt{2l'+1} \, 
 \left( \begin{array}{ccc} l & 2 & l' \\ 0 & 0 & 0 \end{array} \right)  \\
 \bigg\{
 (c_1)^2 \, \left( \begin{array}{ccc} 1 & 1 & 2 \\ 1 & 1 & -2 \end{array} \right) 
 \, \left( \delta_{\xi, \xi'} + i \, \xi \, \delta_{\xi, -\xi'} \right) 
 \, \left( \frac{e^{-i \, 2 \varphi_{E,B}} + \xi \xi' e^{i \, 2 \varphi_{E,B}}}{2} \right) \\
   \bigg[ 
   \left( \begin{array}{ccc} l & 2 & l' \\ \bar{m}_l & -2 & - \bar{m}_l' \end{array} \right) 
 + \xi \xi' \left( \begin{array}{ccc} l & 2 & l' \\ -\bar{m}_l & -2 & \bar{m}_l' \end{array} \right) 
 + \xi' (-1)^{\bar{m}_l'} \, \left( \begin{array}{ccc} l & 2 & l' \\ \bar{m}_l & -2 & \bar{m}_l' \end{array} \right) 
        \bigg] \\
 + 2 \, c_0 \, c_1 \, \left( \begin{array}{ccc} 1 & 1 & 2 \\ 0 & 1 & -1 \end{array} \right) 
  \, \left( \delta_{\xi, \xi'} + i \, \xi \, \delta_{\xi, -\xi'} \right)
   \, \left( \frac{e^{-i \, \varphi_{E,B}} + \xi \xi' e^{i \, \varphi_{E,B}}}{2} \right) \\
  \bigg[ 
   \left( \begin{array}{ccc} l & 2 & l' \\ \bar{m}_l & -1 & -\bar{m}_l' \end{array} \right) 
 - \xi \xi' \left( \begin{array}{ccc} l & 2 & l' \\ -\bar{m}_l & -1 & \bar{m}_l' \end{array} \right) 
 + \xi' (-1)^{\bar{m}_l'}  \, \left( \begin{array}{ccc} l & 2 & l' \\ \bar{m}_l & -1 & \bar{m}_l' \end{array} \right) 
        \bigg] \\        
+ \delta_{\xi,\xi'} \, \delta_{\bar{m}_l,\bar{m}_l'} \, 
   \bigg[
   (c_0)^2 \, \left( \begin{array}{ccc} 1 & 1 & 2 \\ 0 & 0 & 0 \end{array} \right) \,
   - 2 \, (c_1)^2 \, \left( \begin{array}{ccc} 1 & 1 & 2 \\ -1 & 1 & 0 \end{array} \right) 
        \bigg]     
   \bigg[ 
   \left( \begin{array}{ccc} l & 2 & l' \\ \bar{m}_l & 0 & -\bar{m}_l' \end{array} \right)
 +  \left( \begin{array}{ccc} l & 2 & l' \\ \bar{m}_l & 0 & \bar{m}_l' \end{array} \right) \delta_{\bar{m}_l',0} \,  \delta_{\xi',1}
        \bigg]        
\bigg\}  ,
\label{VDDSYM}
\end{multline}
\end{widetext}
with
\begin{eqnarray}
c_{p_i=0,1} = \sqrt{\frac{4 \pi}{3}} \  Y_1^{-p_i}(\theta_{E,B},\varphi_{E,B}=0),
\end{eqnarray}
namely $c_0 = \cos{\theta_{E,B}}$ and $c_1 = \sin{\theta_{E,B}}/\sqrt{2}$.
Equation~\eqref{VVDWSYM} is diagonal in $\xi, \bar{m}_l, l$
and Eq.~\eqref{VHOSYM} is diagonal in $\xi, \bar{m}_l$. Equation~\eqref{VDDSYM}
is diagonal in $\xi$ if $\varphi_{E}=0$ 
and/or $\varphi_{B}=0$, and it reduces in this case to
\begin{widetext}
\begin{multline}
\langle l, \bar{m}_l, \xi | V_{dd}^{e,m} | l', \bar{m}_l', \xi' \rangle =  
 - \frac{\sqrt{30} \, X_{e,m}^2}{r^3} \, \frac{1}{\sqrt{\Delta \, \Delta'}} 
 (-1)^{\bar{m}_l} \, \sqrt{2l+1} \, \sqrt{2l'+1} \, 
 \left( \begin{array}{ccc} l & 2 & l' \\ 0 & 0 & 0 \end{array} \right) \, \delta_{\xi,\xi'} \\
 \bigg\{
 (c_1)^2 \, \left( \begin{array}{ccc} 1 & 1 & 2 \\ 1 & 1 & -2 \end{array} \right) 
  \bigg[ 
   \left( \begin{array}{ccc} l & 2 & l' \\ \bar{m}_l & -2 & - \bar{m}_l' \end{array} \right) 
 + \xi \xi' \left( \begin{array}{ccc} l & 2 & l' \\ -\bar{m}_l & -2 & \bar{m}_l' \end{array} \right) 
 + \xi' (-1)^{\bar{m}_l'} \, \left( \begin{array}{ccc} l & 2 & l' \\ \bar{m}_l & -2 & \bar{m}_l' \end{array} \right) 
        \bigg] \\
 + 2 \, c_0 \, c_1 \, \left( \begin{array}{ccc} 1 & 1 & 2 \\ 0 & 1 & -1 \end{array} \right) 
  \bigg[ 
   \left( \begin{array}{ccc} l & 2 & l' \\ \bar{m}_l & -1 & -\bar{m}_l' \end{array} \right) 
 - \xi \xi' \left( \begin{array}{ccc} l & 2 & l' \\ -\bar{m}_l & -1 & \bar{m}_l' \end{array} \right) 
 + \xi' (-1)^{\bar{m}_l'}  \, \left( \begin{array}{ccc} l & 2 & l' \\ \bar{m}_l & -1 & \bar{m}_l' \end{array} \right) 
        \bigg] \\        
+  \delta_{\bar{m}_l,\bar{m}_l'} \, 
 \bigg[
   (c_0)^2 \, \left( \begin{array}{ccc} 1 & 1 & 2 \\ 0 & 0 & 0 \end{array} \right) \,
   - 2 \, (c_1)^2 \, \left( \begin{array}{ccc} 1 & 1 & 2 \\ -1 & 1 & 0 \end{array} \right) 
        \bigg]      
  \times \bigg[ 
   \left( \begin{array}{ccc} l & 2 & l' \\ \bar{m}_l & 0 & -\bar{m}_l' \end{array} \right)
 +  \left( \begin{array}{ccc} l & 2 & l' \\ \bar{m}_l & 0 & \bar{m}_l' \end{array} \right) \delta_{\bar{m}_l',0} \,  \delta_{\xi',1}
        \bigg]        
\bigg\}  .
\label{VDDSYMZEROPHI}
\end{multline}
\end{widetext}
We can see that we recover a good
quantum number $\xi$ in this formalism which was not the case in the non-symmetrized spherical harmonics basis set.
When $\varphi_{E} = \varphi_{B} \neq 0$, 
or when a unique electric or magnetic field is used 
where $\varphi_{E} \ne 0$ or $\varphi_{B} \ne 0$,
one can always recover the situation in Eq.~\eqref{VDDSYMZEROPHI}
with good quantum numbers
by a proper rotation of the $x$ and $y$ axis 
due to the cylindrical symmetry of the pancakes.
Only in the more general case including both electric and magnetic fields where $\varphi_{E} \neq \varphi_{B}$, one of the dipole-dipole expression is no more diagonal in $\xi$ 
and Eq.~\eqref{VDDSYM} has to be used instead. 
To simplify the study in the following, we will consider the case where $\varphi_{E,B}=0$ when the dipoles are only tilted in the $xOz$ plane. Furthermore we will consider the case where we only apply a unique field (electric in this study).

In general, the quantum number $\bar{m}_l=0$ is 
always automatically associated with
the $\xi=+1$ one 
as it is not defined for the $\xi = -1$ one.
For the $\bar{m}_l > 0$ values one has to use both $\xi= \pm 1$ values.
For the special case of parallel field $\theta_{E,B}=0$, $\bar{m}_l$ is 
a good quantum number, $c_1=0$ and 
Eq.~\eqref{VDDSYM} reduces to
Eq.~\eqref{VDD}. 
Still for the parallel case for $\bar{m}_l \ne 0$, 
the $\xi = -1$ contribution is identical to the $\xi = +1$ one
so that the total contribution is twice the $\xi = +1$ one.
For $\bar{m}_l = 0$, only the $\xi = +1$ contribution is needed.
Therefore, only the value of $\xi = +1$ is needed 
in the parallel field case for all $\bar{m}_l$.
The contribution of the different quantum numbers needed
for different tilted configurations are summarized in Table~\ref{TABLE1}.

\begin{figure*} [t]
\begin{center}
\includegraphics*[width=6cm,keepaspectratio=true,angle=-90]{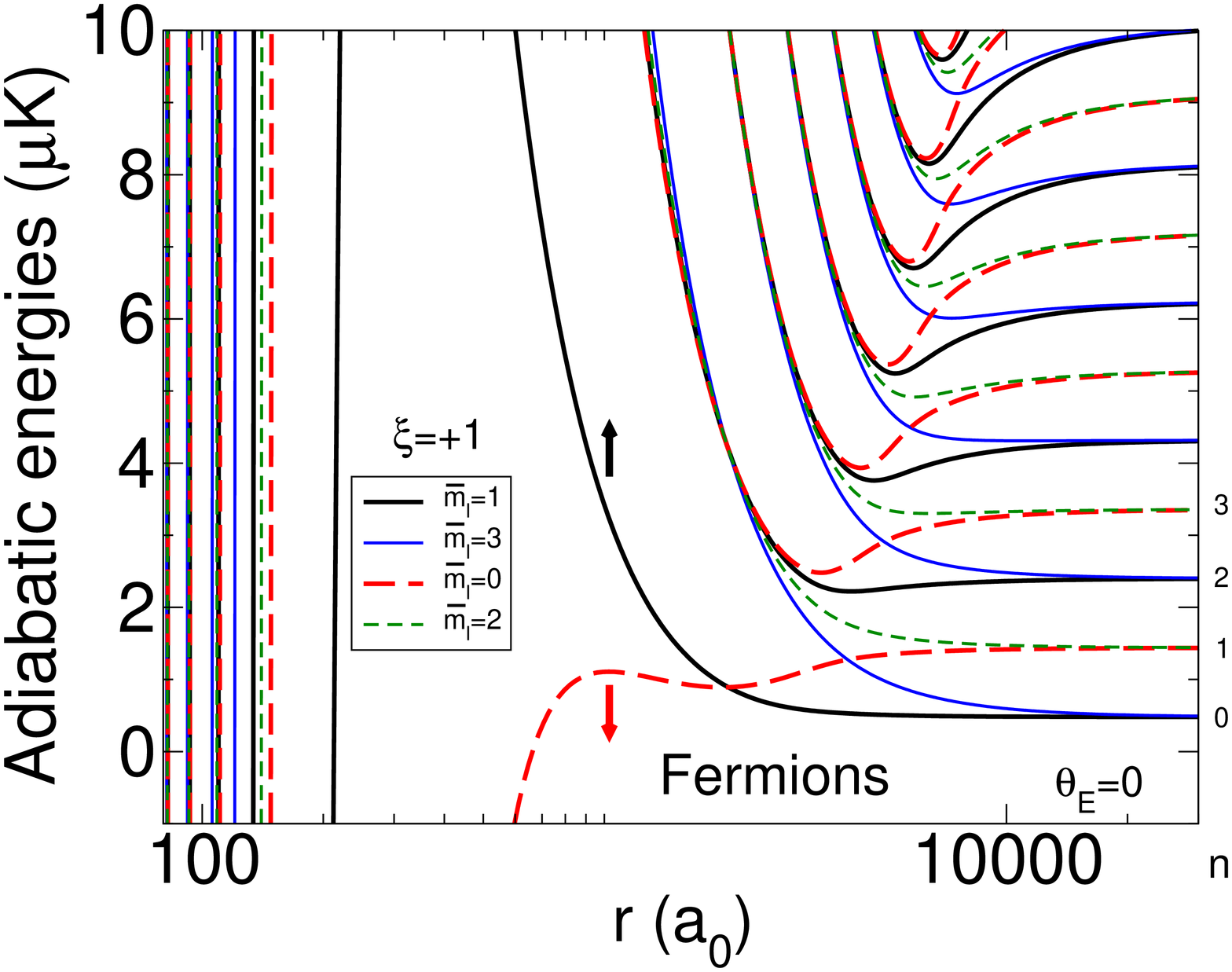} 
\includegraphics*[width=6cm,keepaspectratio=true,angle=-90]{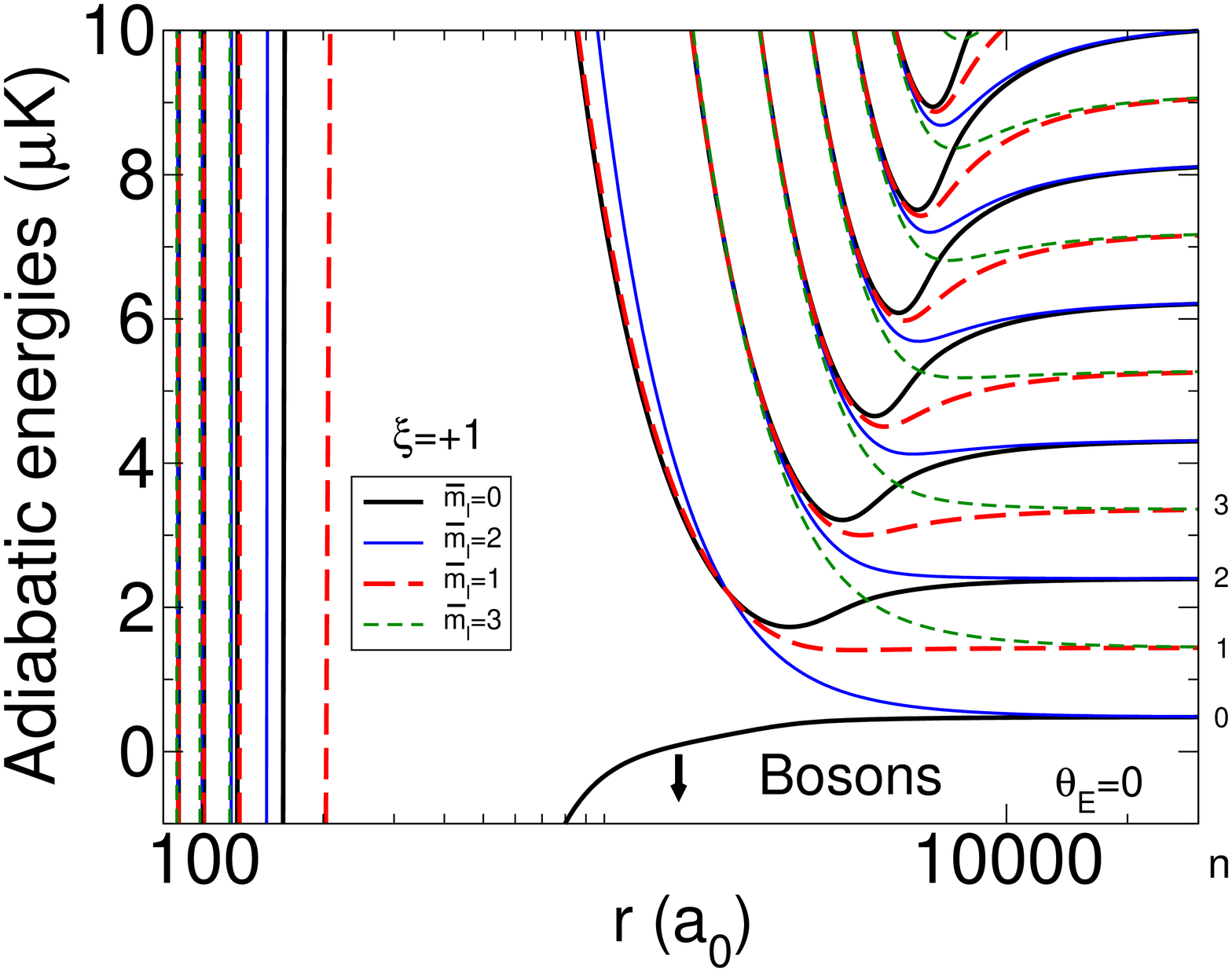}\\
\includegraphics*[width=6cm,keepaspectratio=true,angle=-90]{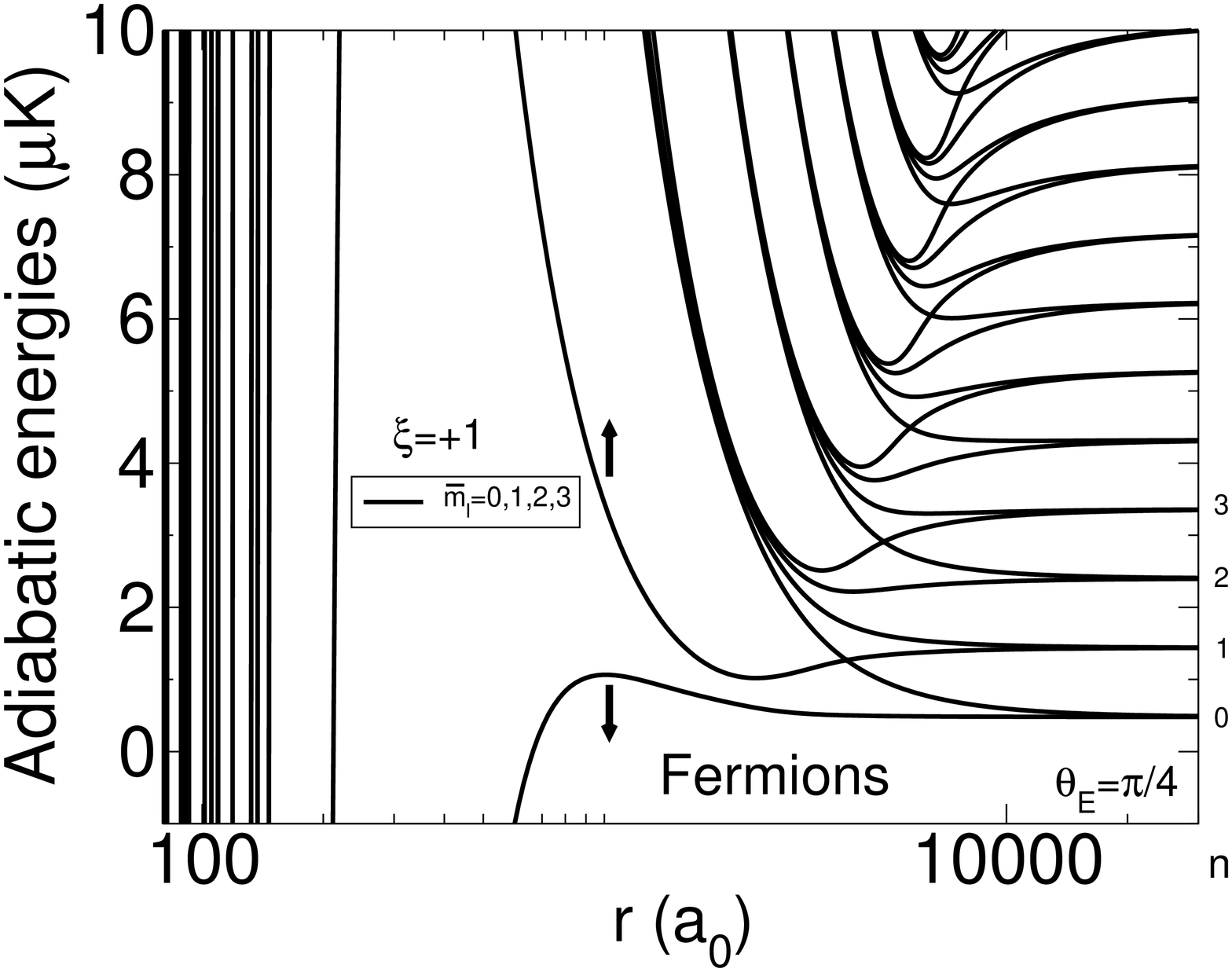}
\includegraphics*[width=6cm,keepaspectratio=true,angle=-90]{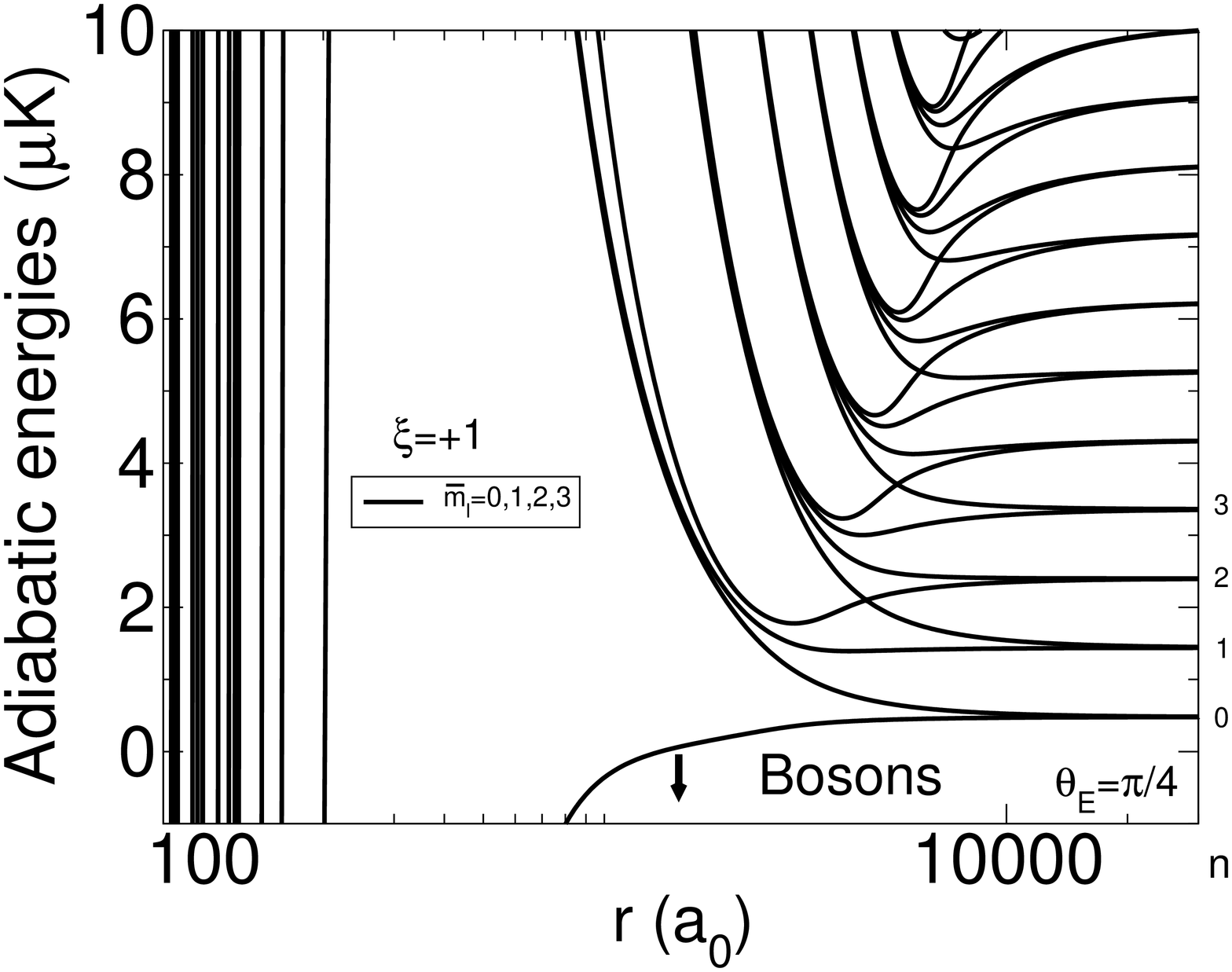} \\
\includegraphics*[width=6cm,keepaspectratio=true,angle=-90]{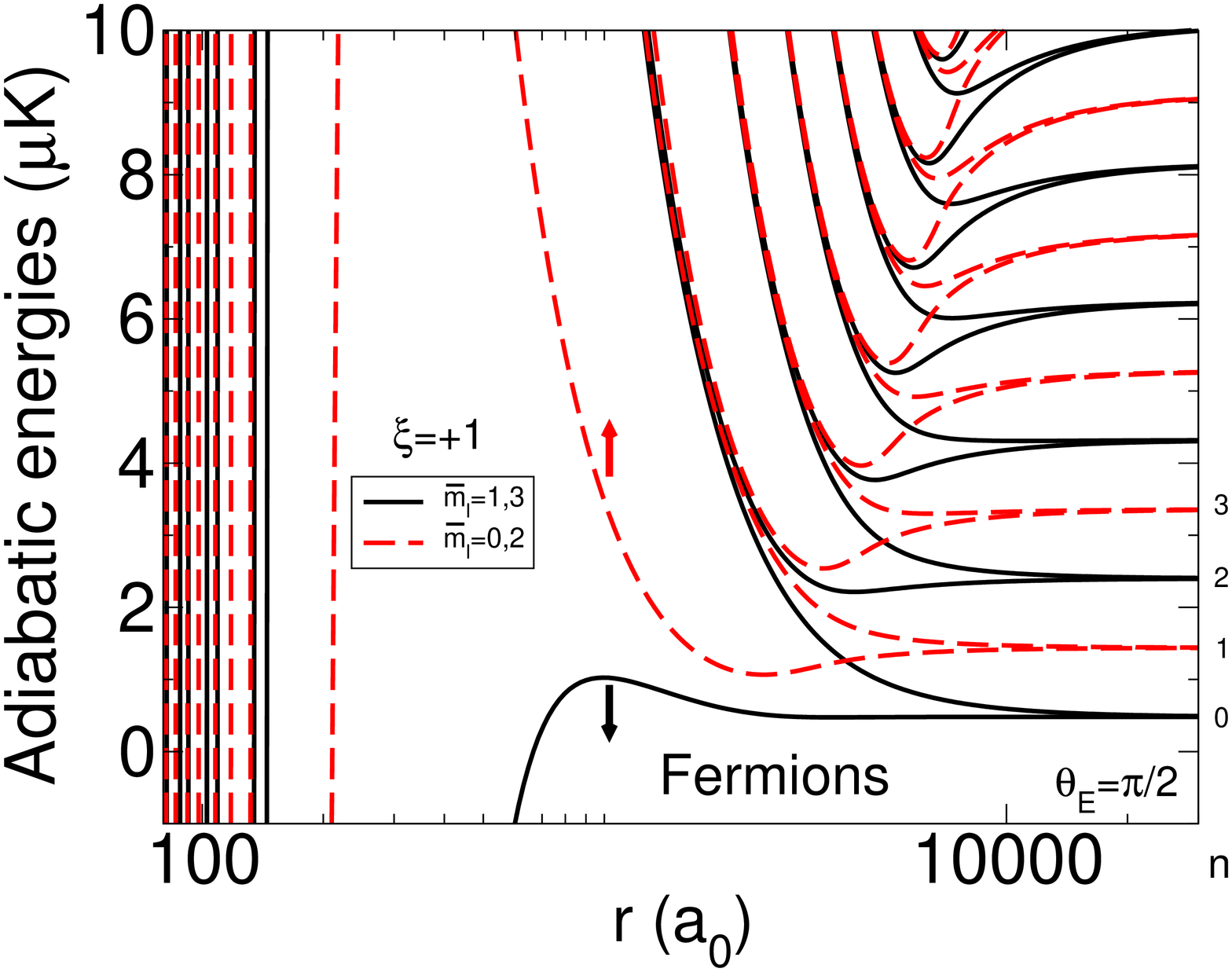} 
\includegraphics*[width=6cm,keepaspectratio=true,angle=-90]{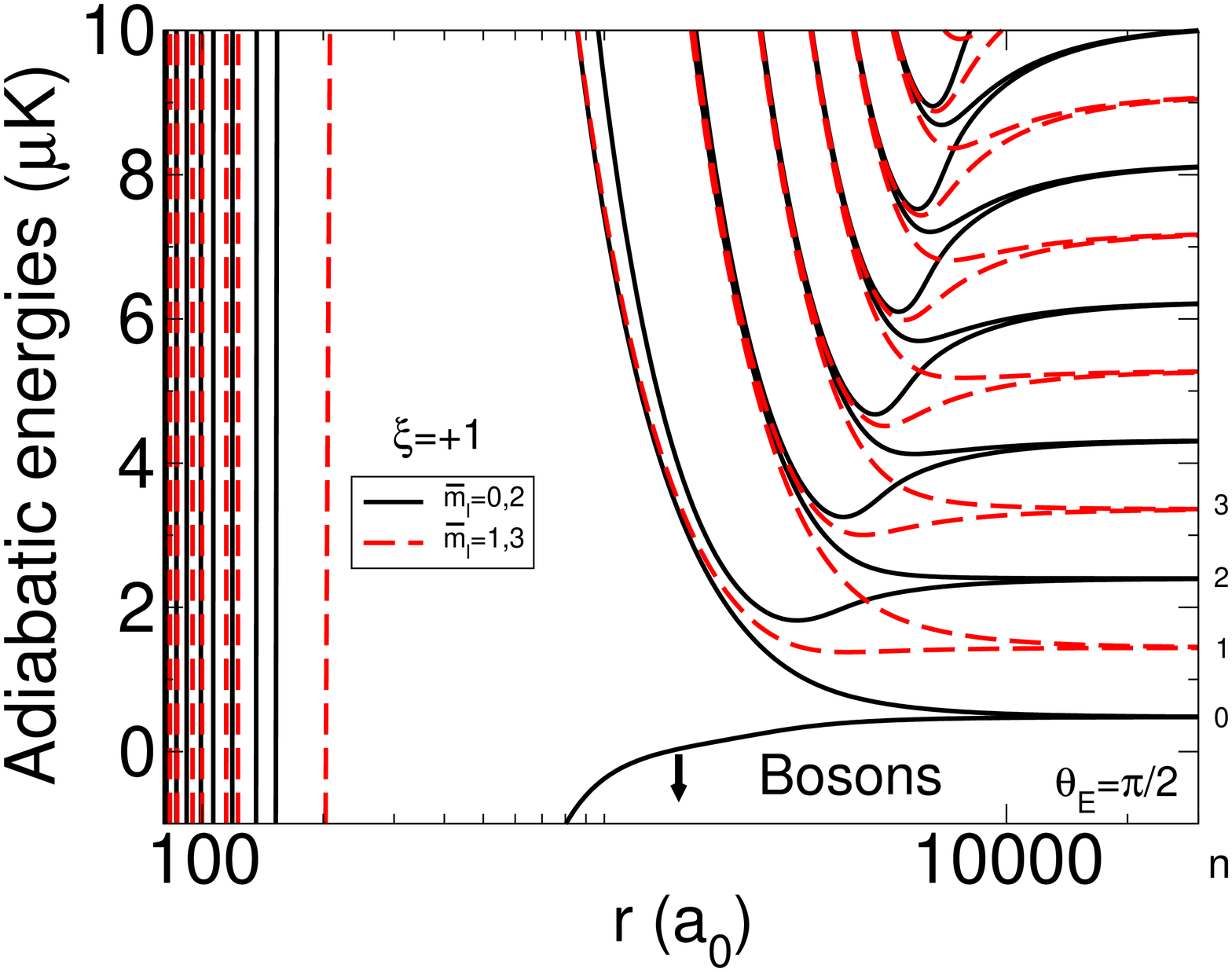} 
\end{center}
\caption{(Color online) Adiabatic energy curves as a function of the intermolecular distance $r$ 
between two 
fermionic $^{40}$K$^{87}$Rb molecules (left panels) and two bosonic $^{41}$K$^{87}$Rb molecules (right panels) for a given induced dipole $d = 0.2$~D 
and tilt angle $\theta_E$. 
The curves of same color are coupled 
among each other. 
Only the curves corresponding to $\xi=+1$ are shown. For the $\xi=-1$ curves, see Appendix A.
Up and down arrows indicate the variation trend of the curves when $d$ increases. Each adiabatic energy curve correlates to a state $n$ of the harmonic oscillator indicated at the right of each panel.}
\label{SPAG-FIG}
\end{figure*}

The suitability of the tesseral harmonics can be qualitatively understood
from their spatial shape. 
If $\varphi = 0$ or $\pi$ which corresponds to the $xOz$ plane 
(see Fig.\ref{DD2DTILT-2-FIG}), $Y_{l,{\bar{m}_l,\xi=-1}}$
vanishes there for any values of $\bar{m}_l$ since it is proportional to 
$\sin{\bar{m}_l \varphi}$ (see Eq.~\eqref{TESSHARM}).
We can therefore associate the $\xi = -1$ quantum number 
with an out-of-plane motion, 
excluding the collision in the $xOz$ plane.
If $\varphi_{E,B}=0$, 
the electric or magnetic field is applied in the $xOz$ plane only.
As the dipoles are exclusively pointing in the $xOz$ plane, 
the collisions of dipoles in the $y$ direction start always 
side-by-side at long-range
since there is no component of the dipole moment in the $y$ direction.
We thus expect that the manifold of the $\xi = -1$ curves
is always more repulsive than its $\xi = +1$ counterpart due to the 
side-by-side, repulsive dipolar approach.

The tesseral harmonics representation of the partial waves is therefore a more appropriate basis set in the general case of collisions in arbitrary tilted electric and/or magnetic fields.
Note that the present formalism is also adapted 
for collisions of particles in free space in crossed electric and magnetic fields, 
where one field (for example the magnetic field) is chosen as the quantization axis and the other 
(for example the electric field) is the tilted one. For example, the study done in
Ref.~\cite{Quemener_PRA_88_012706_2013} using spherical harmonics could 
be adapted using tesseral harmonics.

\section{Results}

In the following, we study the collision of two indistinguishable, electric 
dipolar molecules of KRb in their absolute internal ground state
for which the ``classical dipoles'' 
assumption is a good description. 
We use
$^{40}$K$^{87}$Rb for a fermionic system example
and $^{41}$K$^{87}$Rb for a bosonic one. 
We impose here $\varphi_{E}=0$. 
They possess a permanent electric dipole moment of $d_p = 
0.57$~D~\cite{Ni_S_322_231_2008}. 
The electric field can therefore induce
a dipole moment $d$ up to $d_p$~\cite{Ni_N_464_1324_2010}.
The results presented here will also be similar for collisions
of magnetic dipolar particles with strong magnetic dipole moment, 
tuned by a tilted magnetic field $\vec{B}$, as performed in Ref.~\cite{Frisch_arXiv_1504_04578_2015}.
The confinement in the $z$ direction
is described by a harmonic oscillator of 
frequency $\nu = 20$~kHz which is a typical value employed in experiments. 
We assume that all the particles 
are in the ground state of the harmonic oscillator $n_1=n_2=0$ 
before the collision, which is equivalent to the relative $n=0$ 
quantum number~\cite{Quemener_PRA_83_012705_2011}.
We study those collisions by varying different 
parameters such as the collision energy, the 
tilted angle and the confinement frequency. 
The loss processes are described 
by a full loss condition at short range
given in Ref.~\cite{Wang_NJP_17_035015_2015}.
The full loss condition corresponds to either
a chemical reaction with full probability 
at short range if the system is reactive,
or if not to a possible ``sticky'' rate condition~\cite{Mayle_PRA_87_012709_2013} 
where the two particles stick together for a sufficient
amount of time and a third one has the 
time to destroy the two-body complex 
equivalent to loss of particles.
Although this mechanism has yet 
to be confirmed and observed in experiments,
we include this possibility as well.
We then consider the elastic, inelastic and loss rate 
coefficients for the four lowest values of 
$\bar{m}_l=0,1,2,3$ which are all mixed in a general tilted field
$0 < \theta_E < \pi/2$. In a parallel
field $\theta_E = 0$, none of the $\bar{m}_l$ are mixed and the corresponding rates 
for each $\bar{m}_l$ are summed altogether. 
In a perpendicular field
$\theta_E = \pi/2$, the rates 
are calculated for the even value components $\bar{m}_l=0,2$ 
and the odd value ones $\bar{m}_l=1,3$. 
For the bosonic case, the values of the mixed $l$ are taken 
from $l=0$ to $l=80$ by steps of two, and for the fermionic case,
from $l=1$ to $l=79$ by steps of two.
The fact that we start with molecules in the ground state of the harmonic oscillator
implies that for the special cases $\theta_E=0$ and $\pi/2$, 
$m_l$ should be odd for fermions and even for bosons~\cite{Quemener_PRA_83_012705_2011}.

\subsection{Interactions and adiabatic energy curves}

%
The adiabatic energy curves are shown in Fig.~\ref{SPAG-FIG} 
for the fermionic and the bosonic KRb molecules, 
for an induced dipole moment of $d=0.2$~D. 
If there is a presence of a barrier
relative to a given collision energy
as $r$ decreases and if this barrier increases
when $d$ increases 
(indicated by an arrow pointing upward in
Fig.~\ref{SPAG-FIG}),
we say that the corresponding
curve is protective
against possible short-range loss.
In the absence of a barrier
or if the barrier decreases
(indicated by an arrow pointing downward),
we say that the corresponding
curve is non-protective.
The adiabatic energy curves are 
presented only for the quantum number
$\xi = +1$. 
For the $\xi=-1$ quantum number, 
the curves are equal
to their $\xi = +1$ counterpart in 
the vanishing dipole moment limit, recovering the isotropic character 
of the long-range van der Waals interaction.
For larger $d$, they are more repulsive due to the fact that the 
$\xi=-1$ manifold corresponds to side-by-side dipolar repulsive collisions as mentioned earlier. The $\xi=-1$ curves are shown in Appendix A.

The fermionic case for $\theta_{E}=0$ is shown
in the top left panel of Fig.~\ref{SPAG-FIG}.
The thick black solid lines (resp. thin blue solid lines, 
thick red dashed lines,
thin green dashed lines) represent the (unmixed) values of 
$\bar{m}_l=1$ (resp. $\bar{m}_l=3$, $\bar{m}_l=0$, $\bar{m}_l=2$).
The lowest curve of the $n=0$ harmonic oscillator ground state 
correlates to the protective, 
repulsive side-by-side $\bar{m}_l=1$ curve. 
For indistinguishable fermionic particles, 
we recall that the scattering takes 
place for odd $l$ partial waves. The lowest curve $l=1$ 
features a $p$-wave barrier.
There is an actual crossing between the lowest curves with $\bar{m}_l= 1$ 
and $\bar{m}_l= 0$ since the $\bar{m}_l$ components do not couple to each other.
The middle panel shows the case for $\theta_{E}=\pi/4$.
The black solid lines represent the curves with 
mixed values $\bar{m}_l=0,1,2,3$.
Since all $\bar{m}_l$ are coupled, the actual crossing at $\theta_{E}=0$ 
becomes an avoided crossing. And the lowest $n=0$ curve correlates to a non-protective curve.
Finally, the bottom panel shows the case for $\theta_{E}=\pi/2$.
The black solid lines represent the curves with
mixed values $\bar{m}_l=1,3$ 
and the red dashed lines the curves with mixed values $\bar{m}_l=0,2$. Both series 
of curves do not couple to each other. 
The lowest $n=0$ curve still correlates
to the non-protective curve. 
So for particles in the lowest state,
we can see how the adiabatic energy curve 
changes from a protective character
coming from of a repulsive side-by-side interaction approach
when $\theta_{E}=0$ to a non-protective one
coming from of an attractive head-to-tail interaction approach
when the dipoles are tilted by $\theta_{E}=\pi/2$.

We get similar results in Fig.~\ref{SPAG-FIG} for the bosonic symmetry.
For indistinguishable bosonic particles, the scattering takes 
place in even $l$ partial waves. The lowest $l=0$ curve 
is barrierless, in contrast with the fermionic particles.
Now the lowest $n=0$ curve correlates
to the barrierless, non-protective curve at short range for all $\theta_{E}$ even $\theta_{E}=0$.
As $\theta_{E}$ increases from 0 to $\pi/2$, 
the $\bar{m}_l=2$ curve pushes slightly downwards
the $\bar{m}_l=0$ curve so that the lowest $n=0$ curve get slightly more attractive. 

The behaviour of these adiabatic energy curves 
have direct consequences on the behaviour of the collisional rate coefficients. This is presented below.

\begin{figure*} [t]
\begin{center}
\includegraphics*[width=6cm,keepaspectratio=true,angle=-90]{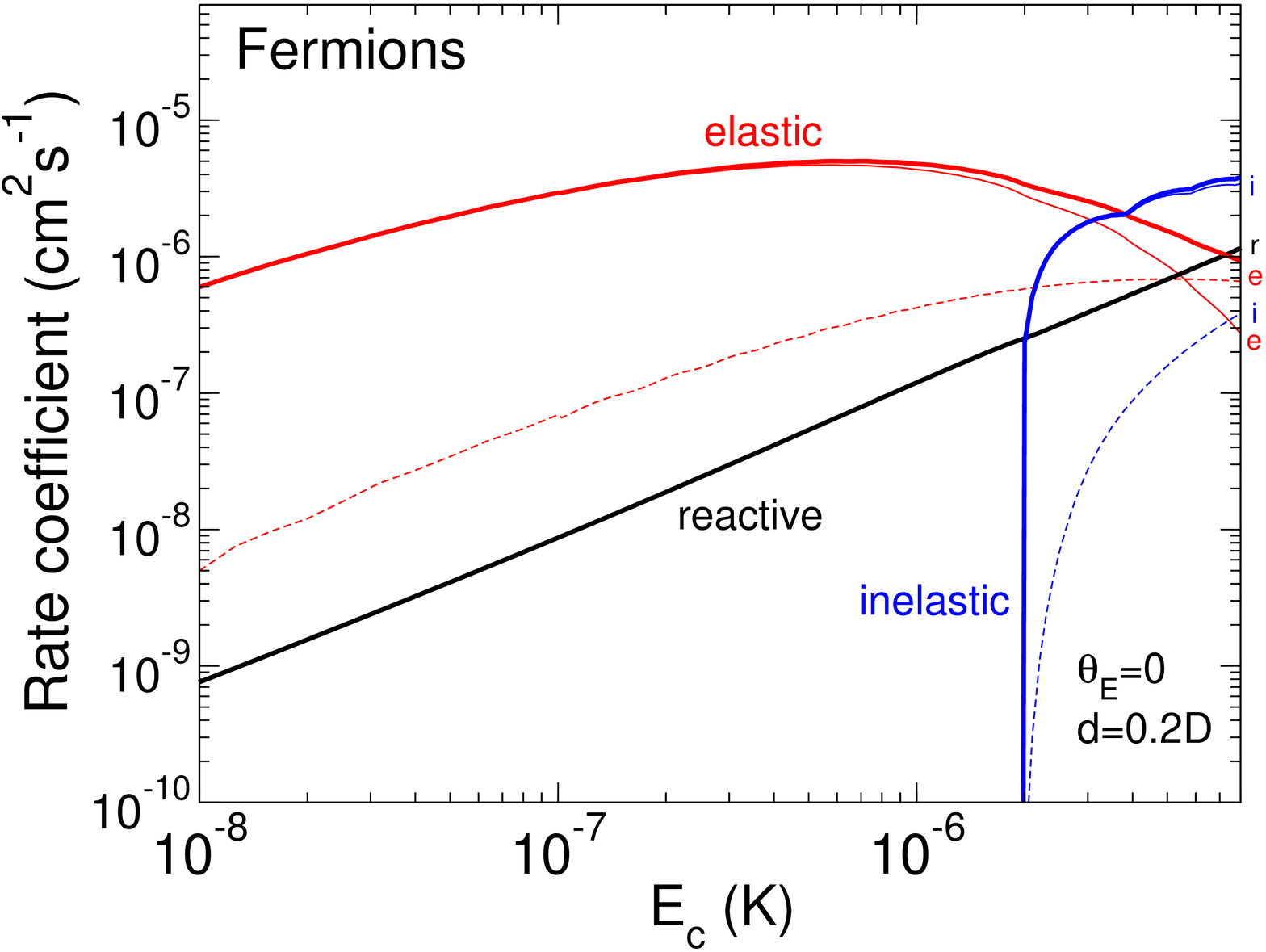} 
\includegraphics*[width=6cm,keepaspectratio=true,angle=-90]{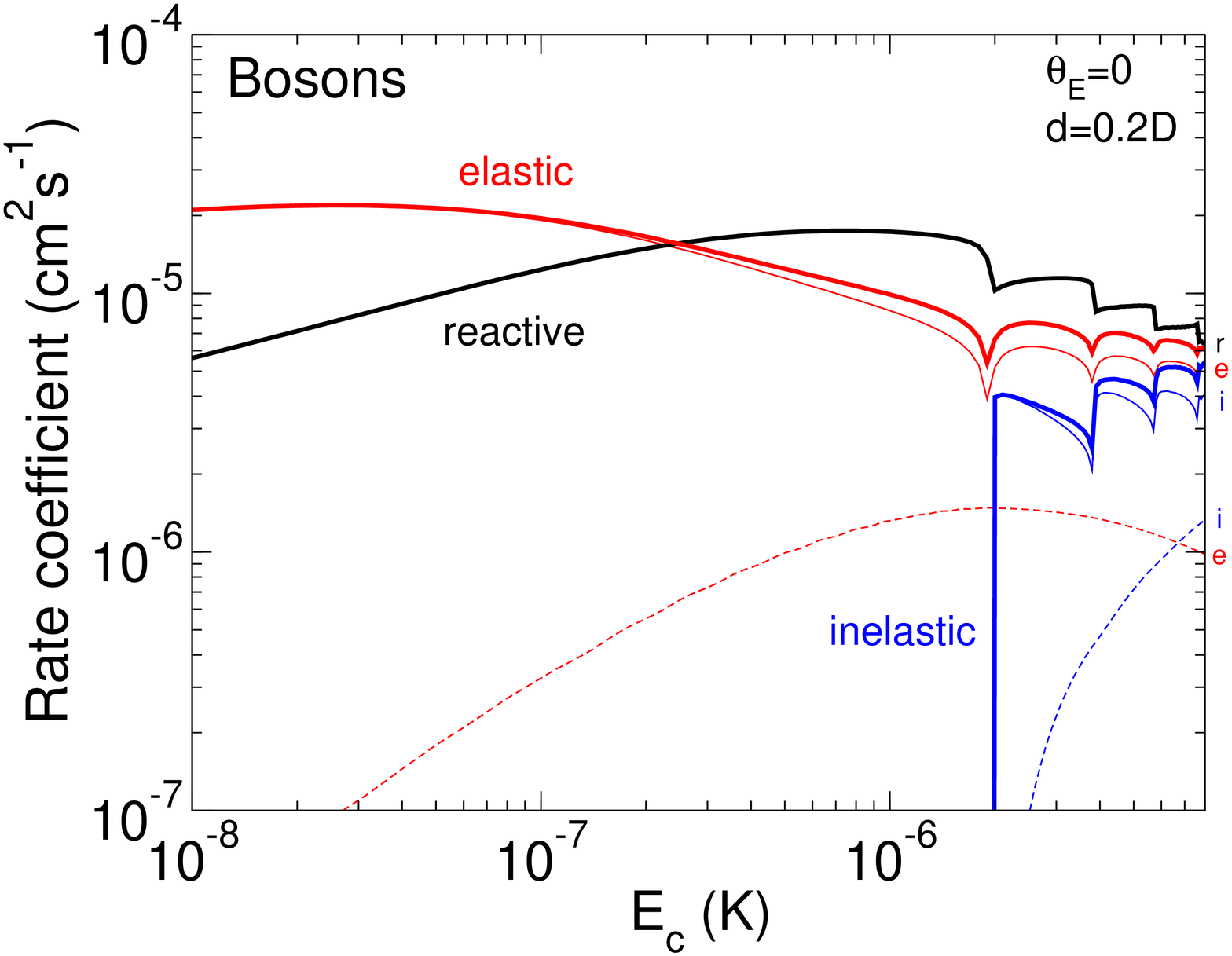} \\
\includegraphics*[width=6cm,keepaspectratio=true,angle=-90]{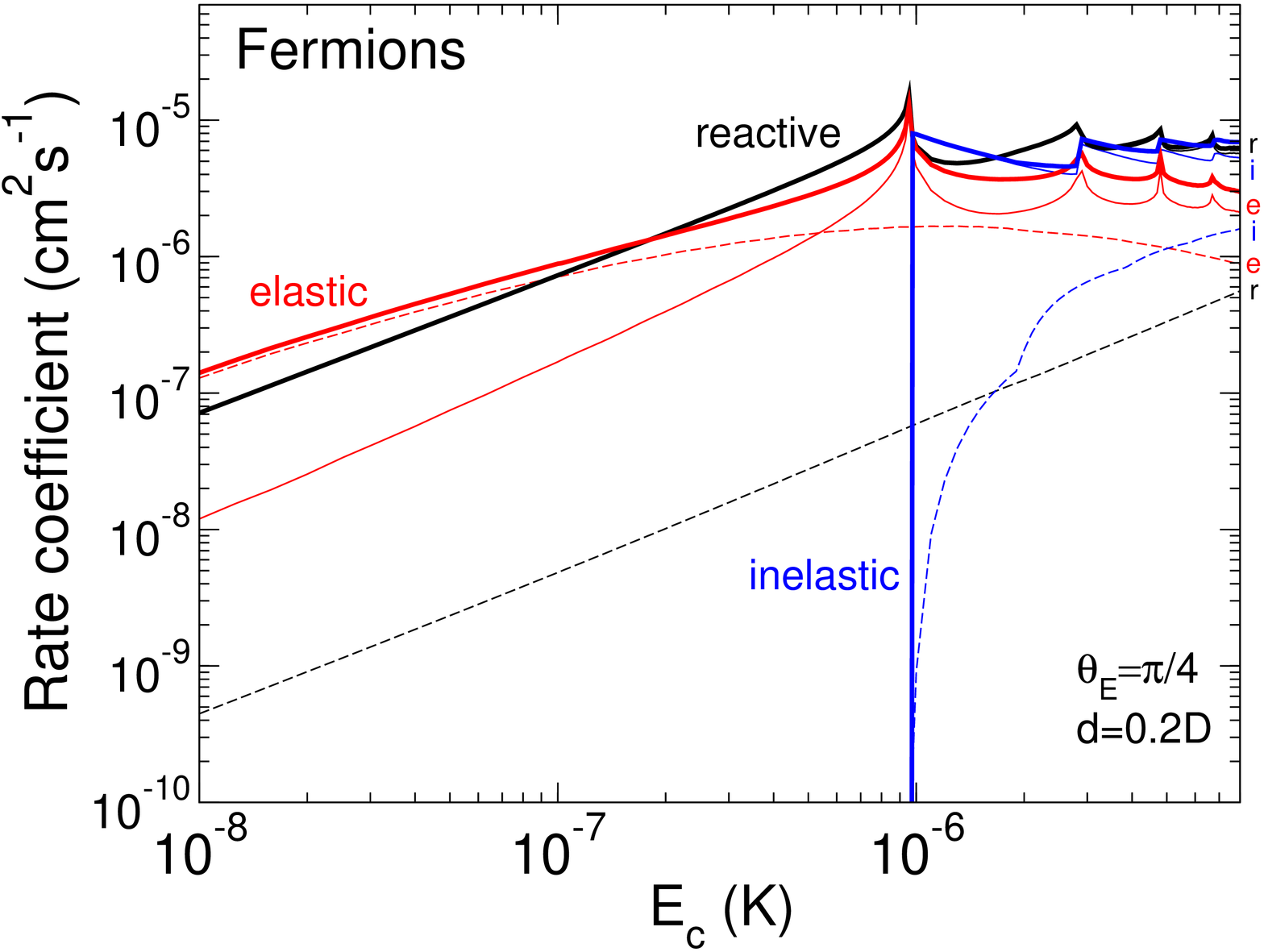} 
\includegraphics*[width=6cm,keepaspectratio=true,angle=-90]{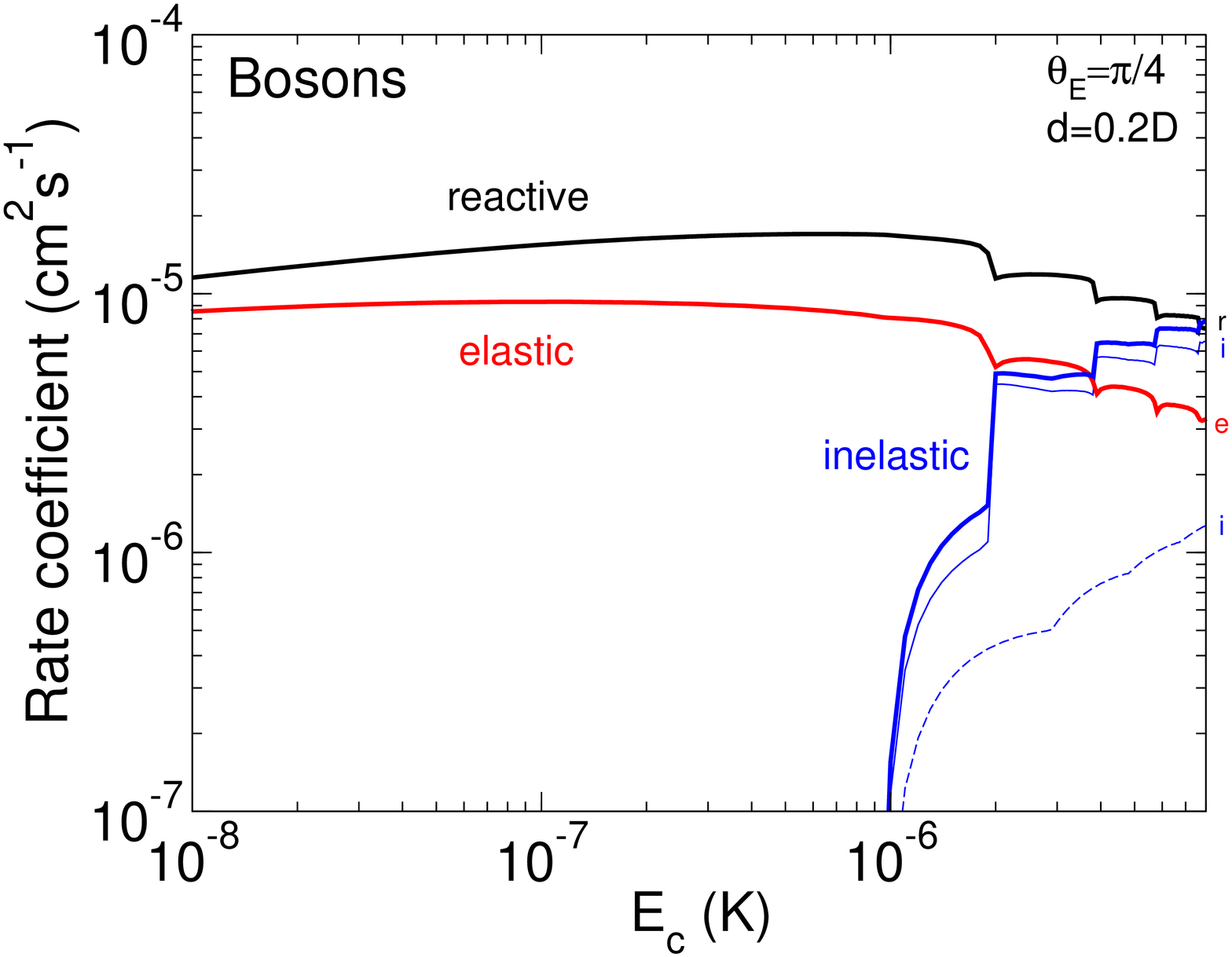} \\
\includegraphics*[width=6cm,keepaspectratio=true,angle=-90]{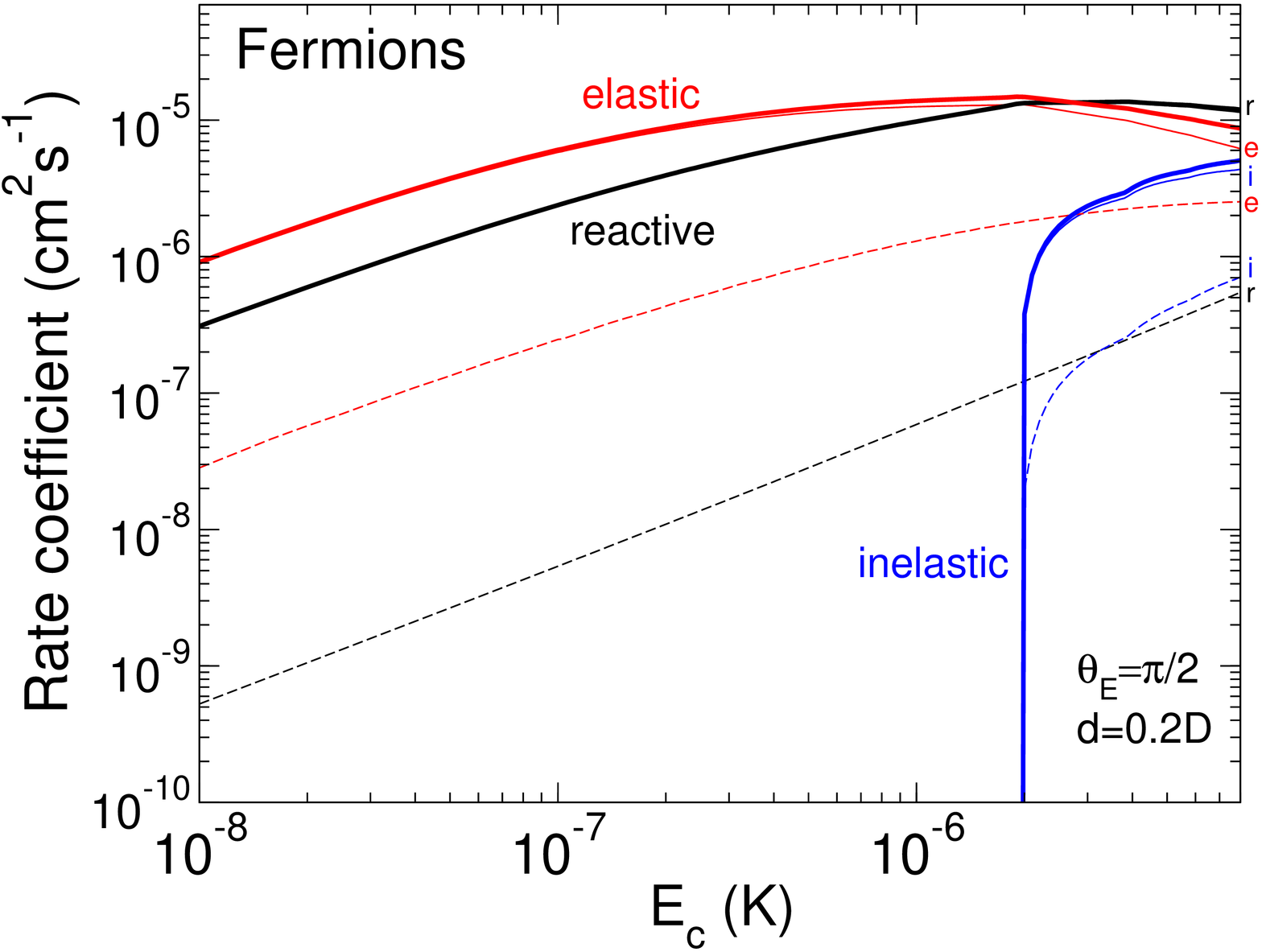}
\includegraphics*[width=6cm,keepaspectratio=true,angle=-90]{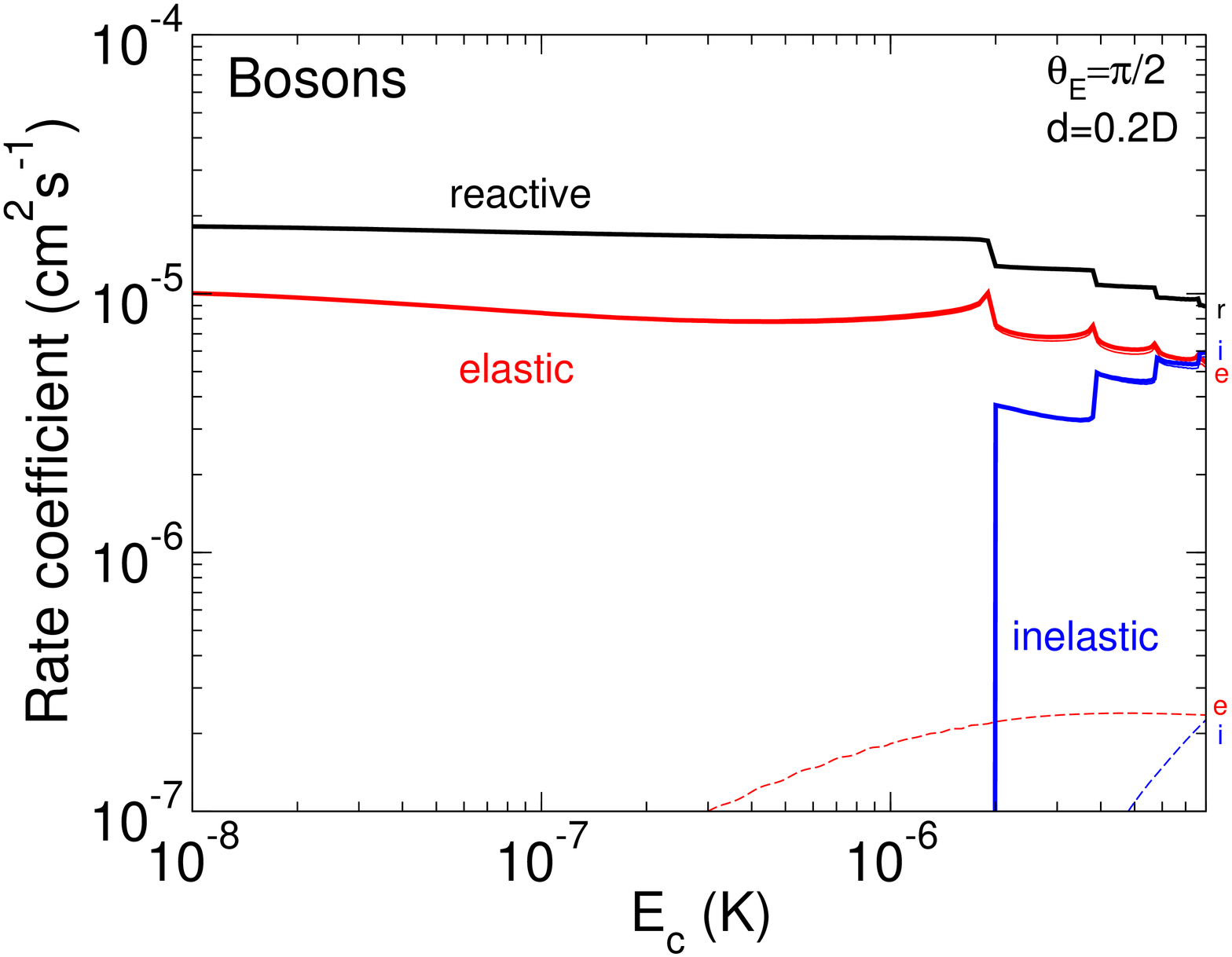} 
\end{center}
\caption{(Color online) Elastic (e), inelastic (i) and reactive (r) rate coefficients (thick lines) for fermionic (left panels) and bosonic (right panels) KRb + KRb collisions as a function of the collision energy at $d$=0.2~D and $\nu = 20$~kHz.
Top panels: $\theta_E = 0$, the thin solid lines are for $\bar{m}_l=1$ and $\xi=\pm1$ ($\bar{m}_l=0$ and $\xi=+1$), the thin dashed lines are for $\bar{m}_l=3$ and $\xi=\pm1$ ($\bar{m}_l=2$ and $\xi=\pm1$) for fermions (bosons).
Middle panels: $\theta_E = \pi/4$, the thin solid lines are for $\xi=+1$, $\bar{m}_l=0,1,2,3$, the thin dashed lines are for $\xi=-1$, $\bar{m}_l=1,2,3$.
Bottom panels: $\theta_E = \pi/2$, the thin solid lines are for $\xi=+1$, $\bar{m}_l=1,3$ ($\bar{m}_l=0,2$), the thin dashed lines are for $\xi=-1$, $\bar{m}_l=1,3$ ($\bar{m}_l=2$)
for fermions (bosons).}
\label{RATEECOLL-FIG}
\end{figure*}

\subsection{Collisions and rate coefficients}

\subsubsection{Rate coefficients versus the collision energy. Threshold laws}

We show in Fig.~\ref{RATEECOLL-FIG} 
the rate coefficients for the 
fermionic and the bosonic case 
as a function of the collision energy for $d=0.2$~D,
starting with two particles in $n=0$ relative motional state.
The top, middle and bottom panels corresponds to 
$\theta_E = 0, \pi/4, \pi/2$ respectively.
The total rates are reported as a thick solid line for each processes.
One can see on Fig.~\ref{SPAG-FIG} that when 
the collision energy is increased from the $n=0$ threshold, 
different harmonic oscillator states 
become energetically open, resulting in sudden peaks in the
inelastic rate coefficients (blue curves).
In addition to the usual elastic rate coefficient (red curves), 
we also have reactive rate coefficients (black curves) 
corresponding to the full loss condition at short range.

For the fermionic particles for the $\theta_E = 0$ case, 
we see that reactive rates are suppressed compared to elastic one at ultralow 
energies. This has already been explained theoretically and experimentally~\cite{Quemener_PRA_81_060701_2010,Quemener_PRA_83_012705_2011,
DeMiranda_NP_7_502_2011}.
As the lowest $n=0$ 
state curve correlates 
to a repulsive protective barrier curve when $r$ 
decreases, 
the probability of the two particles to be close to each other is low, preventing short-range loss to take place.
The inelastic rate coefficient emerges at a collision energy of 
$E_c \sim 2 \, \mu$K corresponding to the $n=2$ threshold opening.
The $n=1$ threshold is not open here at $E_c \sim 1 \, \mu$K 
since the $\bar{m}_l=0$ and 2 (red and green curves)
do not mix with the $\bar{m}_l=1$ and 3 (black and blue curves).
When open above $E_c \sim 2 \, \mu$K, 
the inelastic rates are bigger than the reactive ones and 
can easily amount or overcome the value of the elastic rates.
For $\theta_E = \pi/4$, the reactive rates are bigger than the elastic ones 
at ultralow energies. This is explained again with the corresponding adiabatic
energy curve where the $n=0$ state correlates to the attractive non-protective barrier due to the coupling of the $\bar{m}_l=0$ curve with the $\bar{m}_l=1$ 
one and the presence of the avoided crossing.
The inelastic rate shows up now at $E_c \sim 1 \, \mu$K since the 
$n=1$ threshold is now allowed and coupled with the $n=0$ one.
For $\theta_E = \pi/2$, we recover the head-to-tail collision with large reactive 
rate coefficient. Due to the mixing of $\bar{m}_l=1,3$ only, the  
$n=1$ threshold is not open here too and only the $n=2$ 
opens up at $E_c \sim 2 \, \mu$K.

For the bosonic particles, 
as the collision takes place on a barrierless $l=0$ 
curve, we see that 
the magnitude of the rates 
is bigger than the corresponding one 
for the fermions.
For $\theta_E = 0$, elastic rates are comparable with the reactive ones (either smaller or bigger depending on the collision energy) but there is no 
protection against collisions here. 
As $\theta_E$ increases for $\theta_E = \pi/4$ and $\pi/2$, 
the reactive rate increases as the 
lowest adiabatic energy curve decreases and becomes the dominant rate coefficient.
For the same reasons than those presented 
above for the fermions, 
the inelastic rates show up
as the $n=2$ threshold opens up at $E_c \sim 2 \, \mu$K
for $\theta_E = 0$ and $\theta_E = \pi/2$.
For $\theta_E = \pi/4$, the $n=1$ threshold also opens up at $E_c \sim 1 \, \mu$K.

For all $\theta_E \ne 0$ plots, the contribution of the $\xi=-1$ curves (thin dashed lines) is marginal for an induced dipole moment of $d=0.2$~D. This is due to the large repulsive character of the corresponding adiabatic energy curves which correspond to side-by-side repulsive dipolar collision in the $y$ direction, shown in Appendix A.

For the inelastic collisions, there are striking differences
at the threshold opening of the inelastic state depending on the field angle.
This can be explained by the different behaviour 
of the quantum threshold laws studied in Ref.~\cite{Li_PRL_100_073202_2008}.
In this reference the threshold laws were derived for inelastic relaxation
for $(k_i \to 0) \ll k_f$
where $k_i$ and $k_f$ represent the initial and final 
wavevector. 
Here we consider that the threshold laws for excitation
processes are the same than the relaxation ones 
by replacing $k_i$ by $k_f$ and the initial $m_{l_i}$ 
by the final $m_{l_f}$~\cite{Sadeghpour_JPBAMOP_33_93_2000}.
Then the inelastic relaxation laws given by Eq.~(11) 
of Ref.~\cite{Li_PRL_100_073202_2008}
where $ k_i \ll k_f$, for one or both of $m_{l_i}$ and $m_{l_f}$ equal to zero,
\begin{eqnarray}
\beta_{\text{in. relax.}}^{m_{l_i} \to m_{l_f}} \propto \frac{k_i^{2 m_{l_i}}}{\ln^2{k_i}}
\propto \frac{E_c^{m_{l_i}}}{\ln^2{\sqrt{2 m_\text{red} E_c}}}
\label{WIGLAW1}
\end{eqnarray}
should translate explicitly for the excitation processes to
\begin{eqnarray}
\beta_{\text{in. exc.}}^{m_{l_i} \to m_{l_f}} \propto \frac{k_f^{2 m_{l_f}}}{\ln^2{k_f}}
\propto \frac{(E_c-E_\text{th})^{m_{l_f}}}{\ln^2{\sqrt{2 m_\text{red} (E_c - E_\text{th}) / \hbar^2}}} .
\label{WIGLAW2}
\end{eqnarray}
where $ k_f \ll k_i$, according to the behavior of the different elements 
of Eq.~10 of Ref.~\cite{Li_PRL_100_073202_2008}.
$E_\text{th}$ corresponds to the excitation threshold energy.
For both $m_{l_i}$ and $m_{l_f}$ different than zero, 
the inelastic rate from Ref.~\cite{Li_PRL_100_073202_2008} 
\begin{eqnarray}
\beta_{\text{in. relax.}}^{m_{l_i} \to m_{l_f}} & \propto & k_i^{2 m_{l_i}} \propto E_c^{m_{l_i}} 
\label{WIGLAW3}
\end{eqnarray}
should translate to
\begin{eqnarray}
\beta_{\text{in. exc.}}^{m_{l_i} \to m_{l_f}} & \propto & k_f^{2 m_{l_f}} 
\propto (E_c-E_\text{th})^{m_{l_f}} .
\label{WIGLAW4}
\end{eqnarray}

In our study, 
for the inelastic excitation rate 
$n=0 \to n=2$ at the $E_c \sim 2 \, \mu$K threshold 
for the specific cases $\theta_E = 0$ and $\theta_E = \pi$/2, 
we found a threshold law of 
$\beta_{\text{in. exc.}}^{1 \to 1} \propto (E_c - E_\text{th})$ 
for the fermions
and $\beta_{\text{in. exc.}}^{0 \to 0} \propto \ln^{-2}{\sqrt{2 m_\text{red} (E_c - E_\text{th}) / \hbar^2}}$ 
for the bosons, in agreement with Eq.~\eqref{WIGLAW4} and Eq.~\eqref{WIGLAW2},
where $E_\text{th} = \varepsilon_{n=2}$.
At $E_c \sim 1 \, \mu$K, 
the inelastic excitation rate $n=0 \to n=1$ for $\theta_E = \pi/4$ 
for the fermionic (resp. bosonic) case,
gets the behaviour of the one for the bosonic (fermionic) 
case at $\theta_E = 0$ or $\pi/2$,
with a sharper (rounder) shape.
This is due to a change of parity in $\bar{m}_l$ in the inelastic transition.
For the fermionic (bosonic) case, the inelastic transition $n=0 \to n=1$ 
corresponds to 
$\bar{m}_l=1 \to \bar{m}_l=0$ ($\bar{m}_l=0 \to \bar{m}_l=1$).
For the bosonic case for the $n=0 \to n=1$ inelastic transition, we found a
threshold law of 
$\beta_{\text{in. exc.}}^{0 \to 1} \propto (E_c - E_\text{th}) \, \ln^{-2}{\sqrt{2 m_\text{red} (E_c - E_\text{th}) / \hbar^2}}$ (round shape)
where $E_\text{th} = \varepsilon_\text{n=1}$ now.
For the fermionic case, 
we found a threshold law of 
$\beta_{\text{in. exc.}}^{1 \to 0} \propto \ln^{-2}{\sqrt{2 m_\text{red} (E_c - E_\text{th}) / \hbar^2}}$ (sharp shape).
Again this is in agreement with Eq.~\eqref{WIGLAW2}. 
Finally, for the elastic and reactive collisions,
the quantum threshold laws are found to be 
$\beta_\text{el} \propto E_c^{2}$ and $\beta_\text{re} \propto E_c$ 
for the fermions in agreement with Eq.~(12) and (14) determined in 
Ref.~\cite{Li_PRL_100_073202_2008} for dipolar collisions in quasi-2D.
For the bosons we find 
$\beta_\text{el} \propto \beta_\text{re} \propto \ln^{-2}{\sqrt{2 m_\text{red} E_c / \hbar^2}}$ 
in agreement with Eq.~(9) and (13) of the same reference.

\subsubsection{Sensitivity of the rate coefficients versus the tilted field angle}

\begin{figure} [t]
\begin{center}
\includegraphics*[width=6cm,keepaspectratio=true,angle=-90]{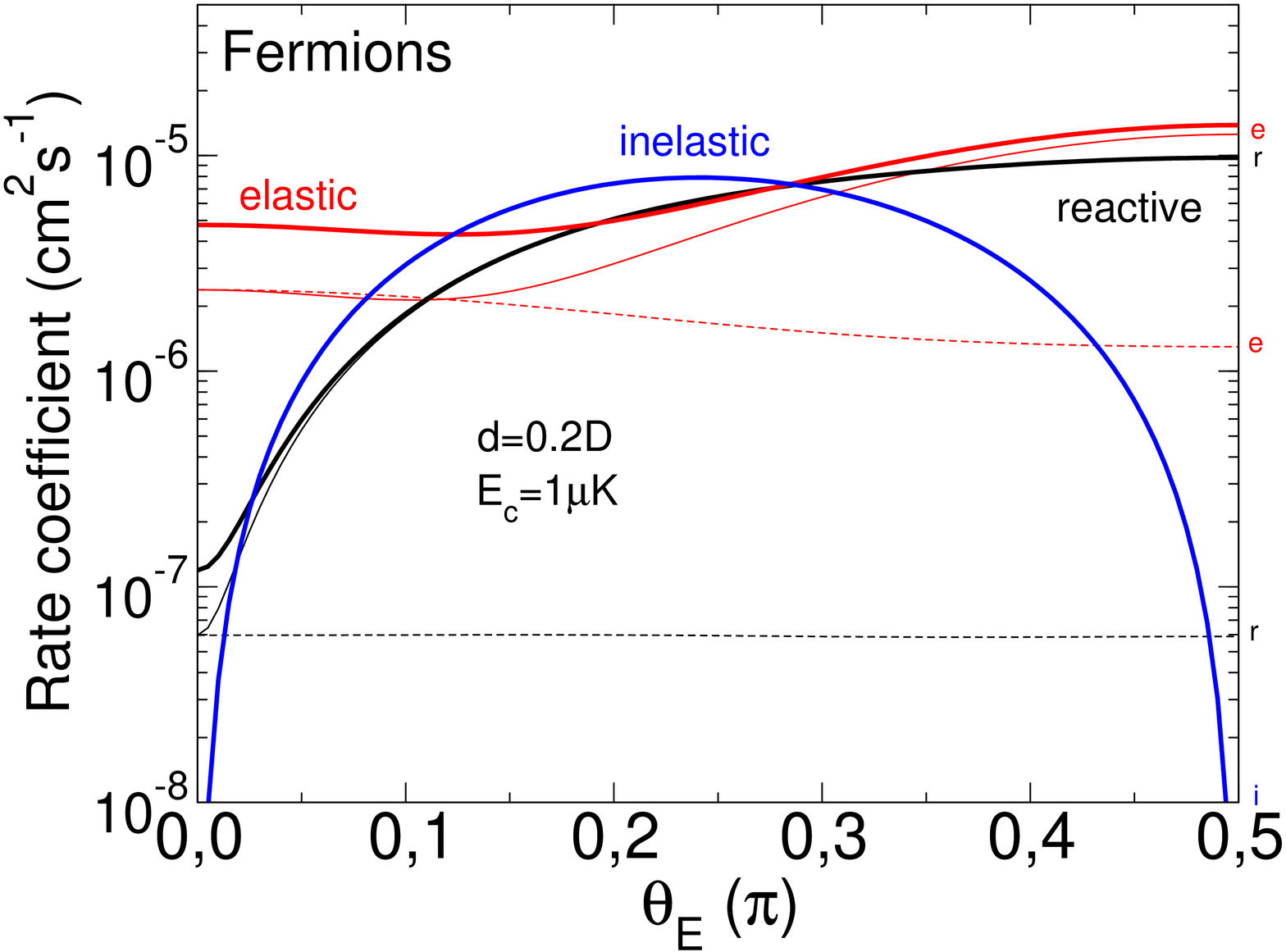} \\
\includegraphics*[width=6cm,keepaspectratio=true,angle=-90]{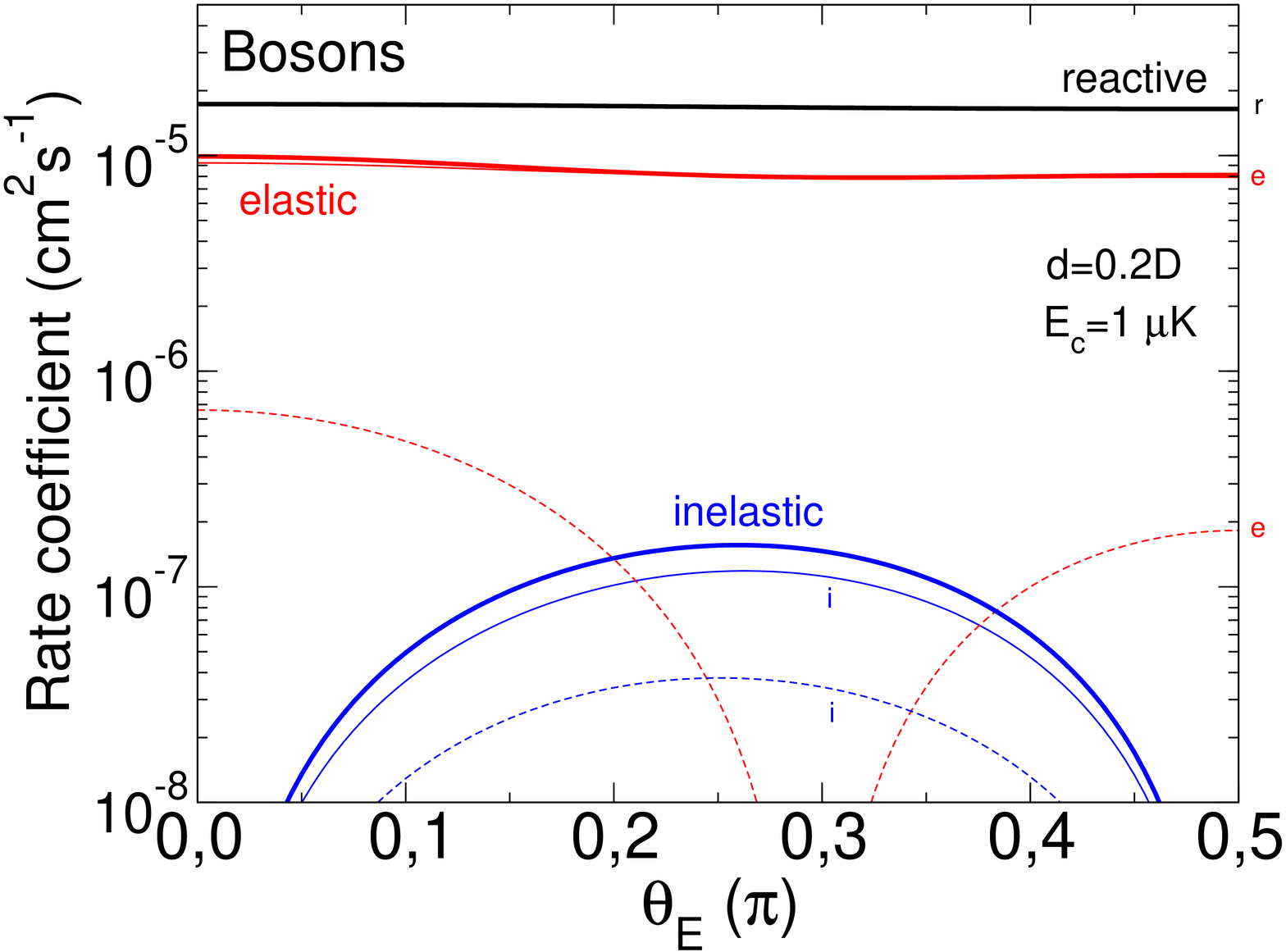}
\end{center}
\caption{(Color online) Elastic (e), inelastic (i) and reactive (r) rate coefficients (thick lines) for fermionic (top panel) and bosonic (bottom panel) KRb + KRb collisions as a function of the tilt angle $\theta_E$, at $E_c$=1~$\mu$K, $d=0.2$~D and $\nu = 20$~kHz. The partial rates are given by thin solid lines ($\xi=+1$, $\bar{m}_l=0,1,2,3$) and thin dashed lines ($\xi=-1$, $\bar{m}_l=1,2,3$).}
\label{RATETEB-FIG}
\end{figure}

The effect of the tilted field on the collision is directly seen in Fig.~\ref{RATETEB-FIG}
for fermions (top panel) and bosons (bottom panel). 
This is plotted at a fixed collision energy of $E_c = 1 \mu$K and induced dipole of $d=0.2$~D.

There is no fermionic and bosonic inelastic rate at $\theta_E=0$ and $\theta_E=\pi/2$ at $E_c = 1 \mu$K. Couplings between $\bar{m}_l=1$, $\bar{m}_l=3$ and $\bar{m}_l=0$, $\bar{m}_l=2$ curves 
are not allowed at these two angles so 
that the transition $n=0 \to n=1$ is forbidden there.
Then the inelastic rates rise from $\theta_E=0$ to $\theta_E=\pi/4$ due to the turning on of the couplings.
At $\theta_E=\pi/4$ there is a maximal coupling between the $\bar{m}_l$
components: 
(i) a maximal coupling for $\Delta \bar{m}_l = 2$
due to the presence of the $\sin^2\theta_E = 1/2$ term in 
Eq.~\eqref{VDDSYMZEROPHI};
(ii) a maximal coupling for $\Delta \bar{m}_l = 1$
due to the presence of the $\cos\theta_E \, \sin\theta_E = 1/2$ term.
Thus the inelastic rates reach their maximal value at $\theta_E=\pi/4$.
Conversely, the couplings turn off as $\theta_E$ passes from $\pi/4$ to 
$\pi/2$, and then the inelastic rates shut off.

The fermionic reactive rate increases 
as we pass from a side-by-side approach $\theta_E = 0$ 
to a head-to-tail one $\theta_E = \pi/2$. 
The continuous transition seen in the rate as $\theta_E$ increases
can be understood qualitativelly 
by the continuous increase of the $\Delta \bar{m}_l = 1$ coupling.
%
%
At $\theta_E=0$ the lowest curve 
of symmetry $\bar{m}_l=1$ 
connects to  the protective barrier 
and leads to a rate $\beta_0$,
while at $\theta_E=\pi/2$
this curve connects to the non-protective barrier 
that leads to a rate $\beta_{\pi/2} \gg \beta_{0} $.
Now, at $\theta_E=0$ the lowest curve 
of symmetry $\bar{m}_l=0$  
connects  to the non-protective barrier, 
the same one that leads to the 
rate $\beta_{\pi/2}$.
Therefore as the coupling between the two symmetries 
increases from $\theta_E = 0$ to $\pi/4$,
the reactive rate is a combination of $\beta_0$ and $\beta_{\pi/2}$
with coefficients that decrease the contribution of $\beta_0$ 
and increase the one of $\beta_{\pi/2}$, hence increasing the total reactive rate.
From $\theta_E = \pi/2$ to $\pi/4$ the reverse argument holds
since now at $\theta_E=\pi/2$ the lowest curve 
of symmetry $\bar{m}_l=1$  ($\bar{m}_l=0$)
connects to the non-protective (protective) barrier.
The reactive rate is again a combination of $\beta_{\pi/2}$ and $\beta_0$.
But as the coupling increases from $\theta_E = \pi/2$ to $\pi/4$,
the coefficient of $\beta_{\pi/2}$ decreases while the one of 
$\beta_0$ increases, hence decreasing the total reactive rates and connecting to the trend
between $\theta_E = 0$ and $\pi/4$.

The sensitivity of the fermionic reactive rate with the field angle 
for the KRb system is quite strong, since 
a small change of $\theta_E = \pi/10$ $(=18^\circ)$ gives rise to an order of magnitude 
increase in the rate. 
Therefore in experiments of fermionic KRb molecules,
it is important for the electric field to be quite 
parallel to the 1D optical lattice 
 confinement axis to avoid additional losses due to slight tilted angles.
 Note however that the range of this strong sensitivity of the rates 
 may vary from one system to another.

In contrast for bosons, there is a very slight dependence of the reactive rate with $\theta_E$. This is due to the slight downward pushing of the $\bar{m}_l=2$ curve to the $\bar{m}_l=0$ ones when $\theta_E$ increases as can be barely seen on Fig.~\ref{SPAG-FIG}. The reactive rate value is anyway larger than the elastic and inelastic one.

Additionally, it is interesting to analyze 
the effect of the $\xi$ component.
For fermions, the $\xi=-1$ rate coefficient turns out to be the same
than the $\xi=+1$ one at $\theta_E = 0$ because
Eq.~\eqref{VDDSYMZEROPHI} reduces to Eq.~\eqref{VDD}
which is an expression independent of the $\xi$ number.
When departing from the angle $\theta_E=0$, 
the $\xi=+1$ rate coefficients are dominant
compared to the $\xi=-1$ ones.
As a good approximation, one can neglect the 
contribution of the latter for large values of $\theta_E$
(note that as $d$ increases, this is less true though, see comment in Appendix A).
For bosons at $\theta_E=0$, the $\xi=+1$ term contains the $\bar{m}_l=0$ component 
responsible for a barierless collision
while intrinsically the $\xi=-1$ one does not. Therefore the latter case does not yield a higher 
rate than the $\xi=+1$ case, then it is a good approximation 
to neglect the $\xi=-1$ component for bosons for all angles $\theta_E$.

\subsubsection{Rate coefficients versus the confinement}

\begin{figure} [t]
\begin{center}
\includegraphics*[width=6cm,keepaspectratio=true,angle=-90]{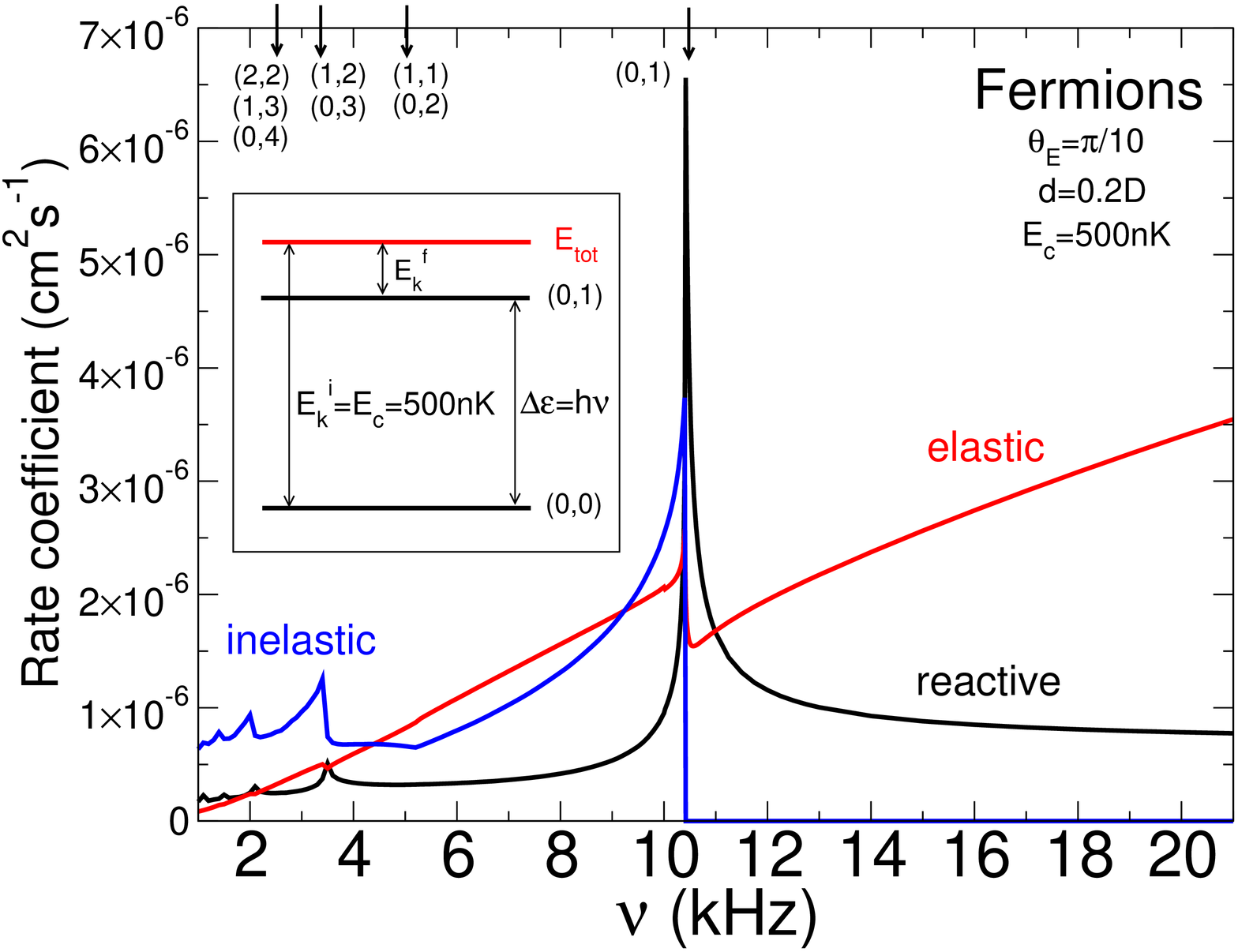} \\ 
\includegraphics*[width=6cm,keepaspectratio=true,angle=-90]{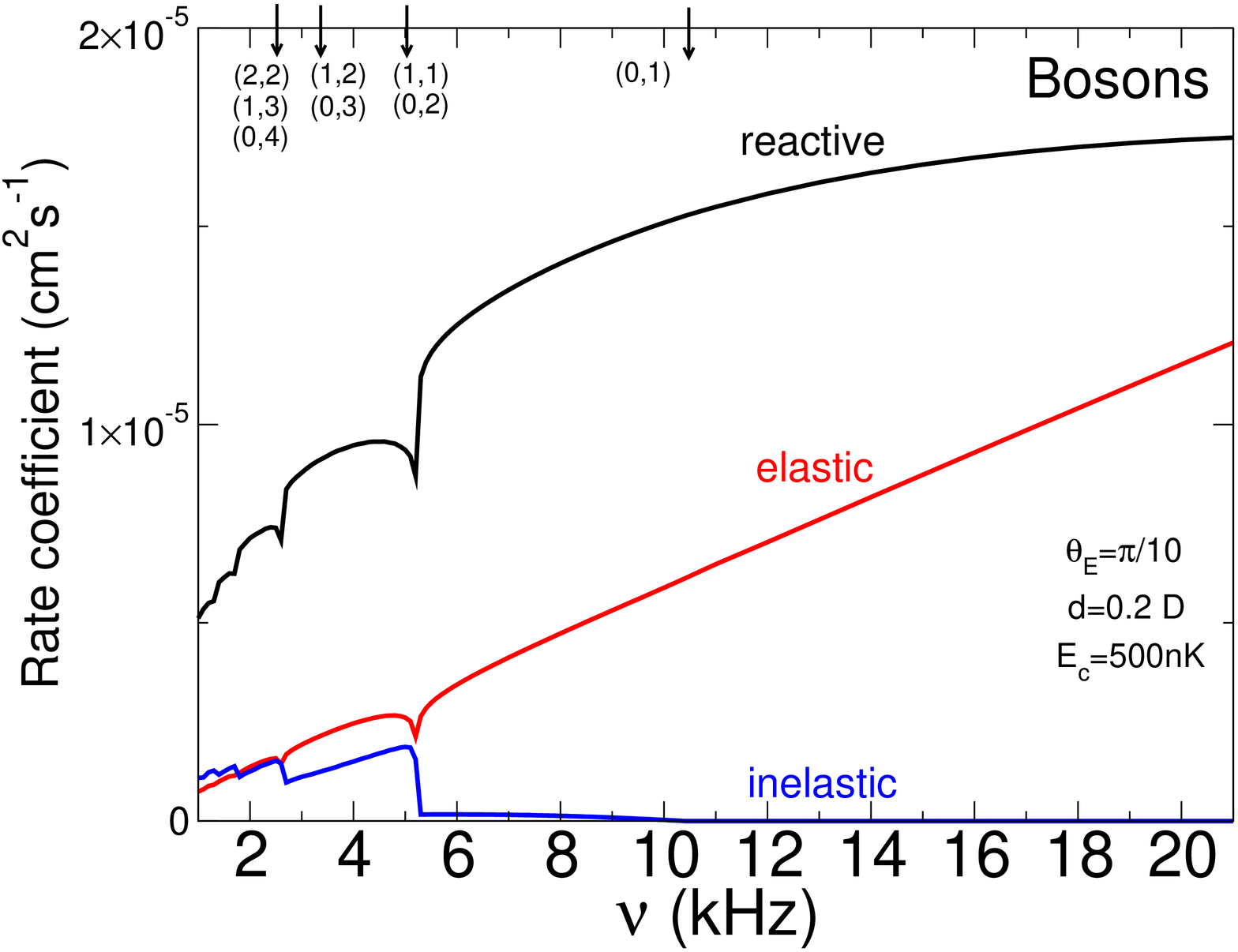}
\end{center}
\caption{(Color online) Elastic, inelastic and reactive rate coefficients for fermionic (top panel) and bosonic (bottom panel) KRb + KRb collisions as a function of the tilt angle $\theta_E $, at E$_c$=500 nK, d=0.2~D and $\nu = 20$~kHz. The numbers in the brackets represent the harmonic states $(n_1,n_2)$ of particles 1 and 2 and the arrows the corresponding energy thresholds. The inset in the top panel sketches the excitation inelastic process $(0,0) \to (0,1)$ (see text for details).}
\label{RATENU-FIG}
\end{figure}

The effect of the confinement strength is shown in Fig.~\ref{RATENU-FIG} for a fixed collision energy of $E_c = 500$~nK, induced dipole of $d=0.2$~D and field angle $\theta_E = \pi/10$ ($18^\circ$). 
At $\nu = 20$~kHz, the first threshold $n_1=0,n_2=1$
is located at an energy of $\sim 1 \mu$K above the energy of the initial state $(n_1=0,n_2=0)$.
Therefore a collision energy of 500 nK is not sufficient to open up inelastic collisions
and there is no inelastic rate.
When the confinement decreases, there is a given $\nu$ for which the first inelastic transition becomes open, when the first threshold $(n_1=0,n_2=1)$ energy amounts the value of the collision energy. This is the case here at $\nu = 10.4$~kHz. These different values of 
$\nu$ are indicated with an arrow along with the different threshold openings $(n_1,n_2)$.
At $\nu = 10.4$~kHz, the first inelastic threshold opening (first arrow from the right) 
for the fermions is quite strong, recalling the sharp one seen in 
Fig.~\ref{RATEECOLL-FIG} for $\theta_E = \pi/4$. The first one for the bosons is rather 
weak and smooth as also seen in Fig.~\ref{RATEECOLL-FIG}. 
The second opening (second arrow from the right) is now smooth for the fermions and 
sharp for the bosons. And so forth, the successive openings alternate 
between sharp and smooth patterns. As explained earlier, the smooth openings correspond to a $\bar{m}_l=0 \to \bar{m}_l=odd$ transition
while the sharp openings correspond to a $\bar{m}_l=1 \to \bar{m}_l=even$ transition.

An interesting feature is seen for the fermionic system. At $\nu = 9$~kHz for example,
the inelastic rate $(n_1=0,n_2=0) \to (n_1=0,n_2=1)$ reaches the value of the elastic rate ($\sim 2 \ 10^{-6}$~cm$^2$~s$^{-1}$) while the reactive rate is about 3.3 times smaller.
The inelastic process here is an excitation inelastic process (see the inset of Fig.~\ref{RATENU-FIG}). 
The total energy is set before the collision by $E_{tot} = \varepsilon_0 + \varepsilon_0 + E_c = 0.932 \mu$K.
We also have $\varepsilon_0 = 0.216 \mu$K and $\varepsilon_1 = 0.648 \mu$K. 
After the collision, the total energy is given by 
$E_{tot} = \varepsilon_0 + \varepsilon_1 + E_k^f$
where $E_k^f$ is the finall kinetic energy. 
The conservation of the total energy
implies $E_k^f = E_c - (\varepsilon_1 - \varepsilon_0) = E_c - h \nu = 500$ nK$ - 432$ nK$ = 68$ nK. This means that for an excitation inelastic process, the particles after the collision have a smaller kinetic energy than before the collision and therefore slowed down by this mechanism.
Loss of molecules can happen but according
to the rates, for each loss of one pair, three pairs get slowed down from 500 nK to 68 nK. 
Note that the final kinetic energy depends on the choice of the frequency trap (see inset): its value is even smaller when the frequency is closer to the frequency that shuts off the inelastic transition, here $\nu = 10.4$ kHz. Therefore, particles can be slowed down to arbitrary small kinetic values, as far as the reactive rates remain smaller than the inelastic ones.
Of course the reverse relaxation inelastic process $(n_1=0,n_2=1) \to (n_1=0,n_2=0)$ can also happen after the particles have been excited, restoring back the 500 nK kinetic energy to the particles. But this can be prevented by tilting the electric field back to the parallel case $\theta_E=0$. In such cases, the inelastic transition is forbidden as discussed previously, leaving the particles in the harmonic states 0 and 1 with small kinetic energy. Another possibility is to find a way to remove directly the particles in the harmonic state 1 leaving only the one in the ground state 0 with small kinetic energy.

\section{Conclusion}

By implementing tesseral harmonics in place of spherical harmonics in the collisional formalism of two ultracold tilted dipolar particles in confined space, we showed that we can recover a good quantum number $\xi$. This separates the overall problem into two sub-problems of smaller size even when a field is tilted. Inelastic and reactive rates show dramatical changes in a tilted field. This is due to additional couplings between $\bar{m}_l$ components
when a tilt is applied. 
We also showed that fermionic dipolar particles 
can lose kinetic energy and slow down 
due to favourable 
trap excitation inelastic collision 
under an appropriate confinement strength of the 1D lattice in a tilted field.
Future works will investigate whether this mechanism
can be efficient to even further cool down the particles in such configuration by taking into account the initial kinetic energy distribution of the particles for a given temperature and their rethermalization due to the elastic colliisons 
during this process.

\section*{Acknowledgments}

We acknowledge the financial support of the COPOMOL project (\# ANR-13-IS04-0004-01) from Agence Nationale de la Recherche, the project Attractivit\'e 2014 from Universit\'e Paris-Sud and the project SPECORYD2 (Contract No. 2012-049T) from Triangle de la Physique. \\

\section*{Appendix A: $\xi=\pm1$ adiabatic energy curves}

\begin{figure}[h]
\begin{center}
\includegraphics*[width=6cm,keepaspectratio=true,angle=-90]{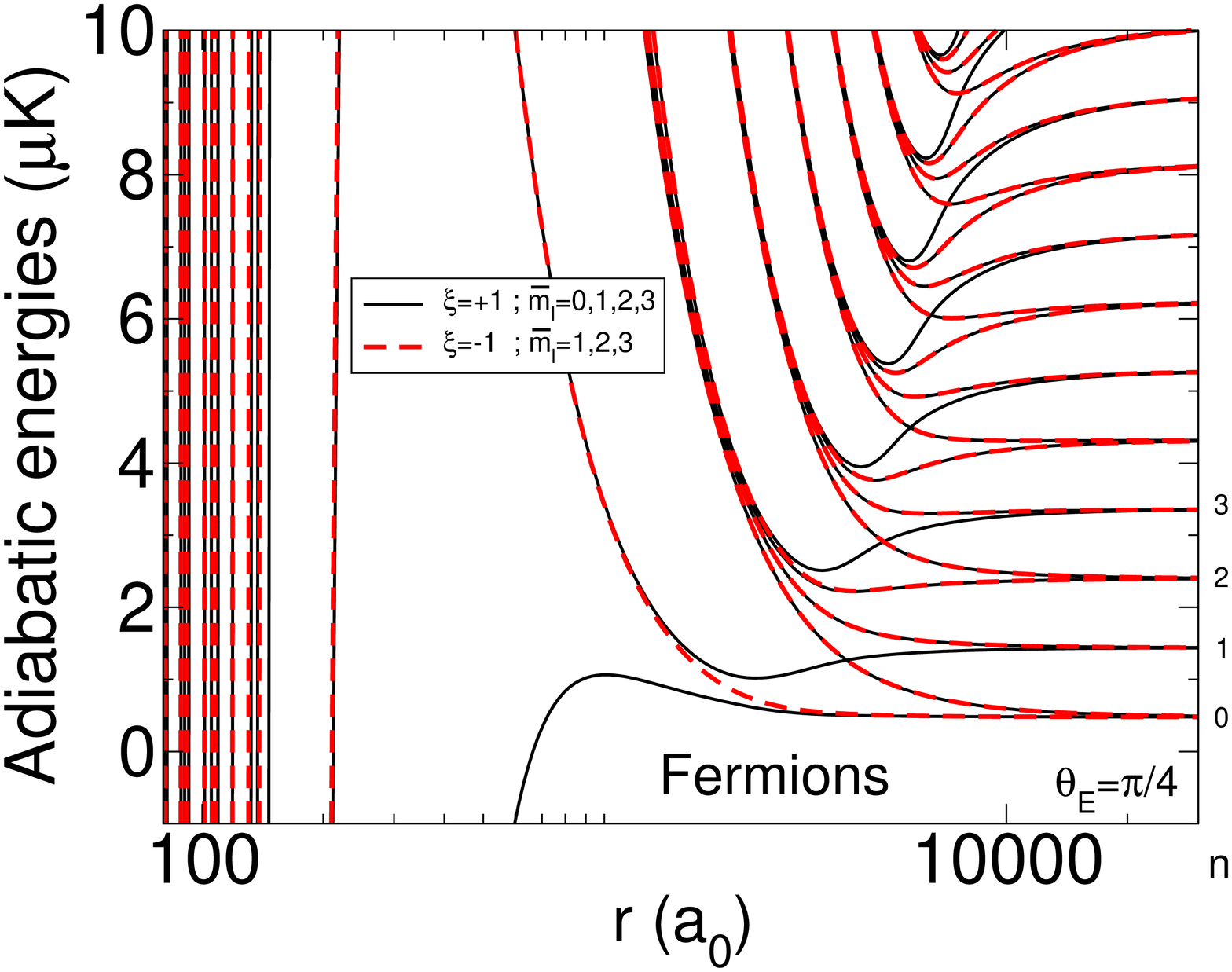} \\
\includegraphics*[width=6cm,keepaspectratio=true,angle=-90]{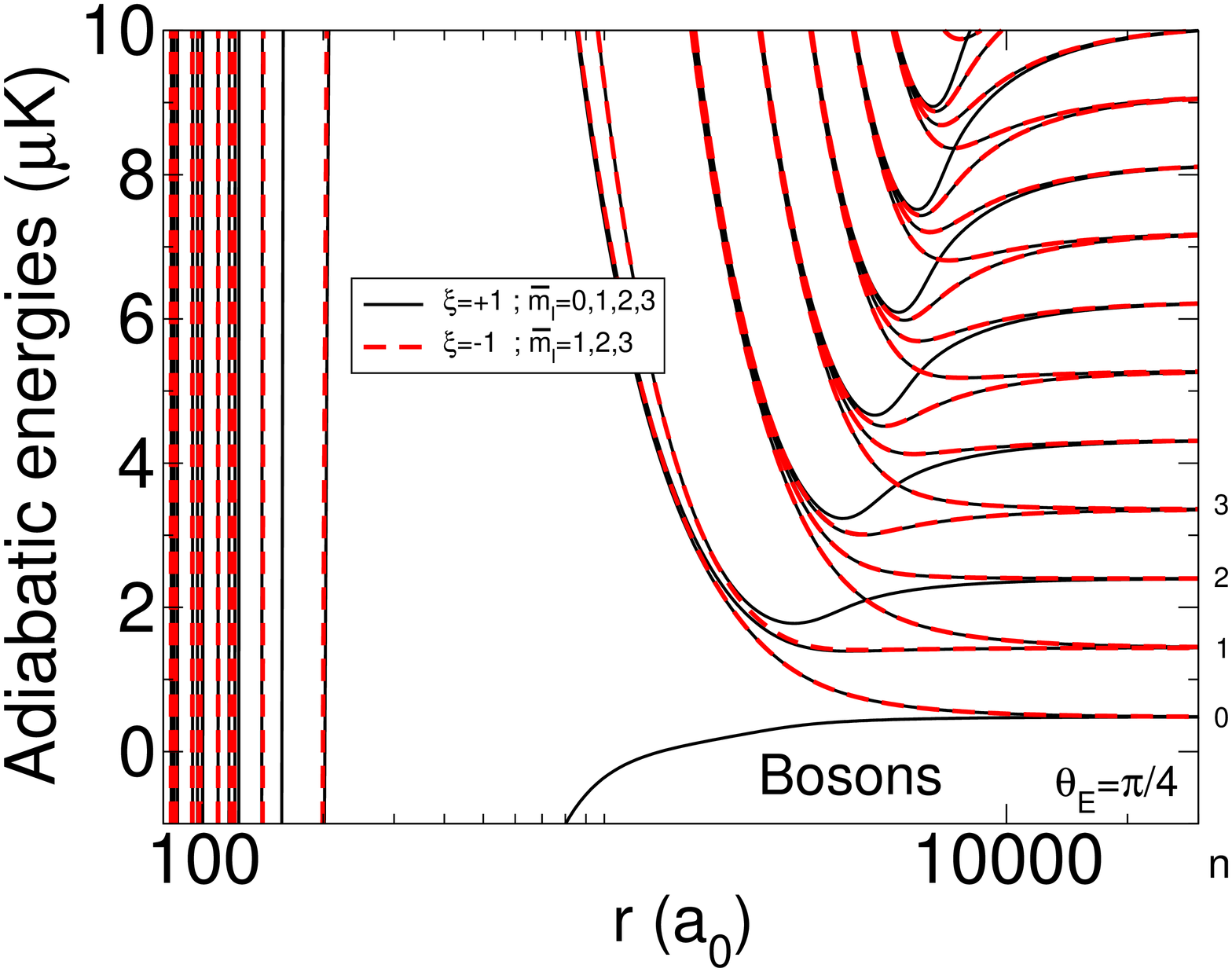}
\end{center}
\caption{(Color online) Adiabatic energy curves as a function of the intermolecular distance $r$ 
for $d=0.2$~D, $\theta_E = \pi/4$ for the $\xi=+1$ manifold (black solid lines) and the $\xi=-1$ manifold (red dashed lines). Fermions: top panel, bosons: bottom panel.
Each adiabatic energy curve correlates to a state $n$ of the harmonic oscillator.}
\label{SPAG-XI=M1-FIG}
\end{figure}

Fig.~\ref{SPAG-XI=M1-FIG} presents the adiabatic energy curves for $d=0.2$~D and $\theta_E = \pi/4$ for 
both the fermionic and bosonic system and both quantum numbers $\xi = \pm 1$. One can see that the 
$\xi=-1$ curves always corresponds to more repulsive curves than $\xi=+1$ ones. This is due to the 
fact that $\bar{m}_l \ge 1$ for $\xi=-1$ and it correlates to a $l \ge 1$ curve with a large centrifugal barrier. At weak 
$d$ the $\xi=-1$ manifold does not play an important role in the collision. 
Note that at large $d$, the $l \ge 1$ curve can turn attractive again
and then the $\xi = -1$ manifold can start to play a role in the collision~\cite{Quemener_PRA_83_012705_2011}.

\bibliography{../../BIBLIOGRAPHY/bibliography.bib}
   
\end{document}